\documentclass{ieeeaccess}
\usepackage{cite}
\usepackage{amsmath,amssymb,amsfonts}
\usepackage{algorithmic}
\usepackage{graphicx}
\usepackage{textcomp}

\usepackage[nolist]{acronym}
\usepackage{caption}
\usepackage{subcaption}
\usepackage{bm}	
\usepackage{booktabs}	
\usepackage{multirow} 
\usepackage{colortbl}

\def\BibTeX{{\rm B\kern-.05em{\sc i\kern-.025em b}\kern-.08em
    T\kern-.1667em\lower.7ex\hbox{E}\kern-.125emX}}

\begin{document}

\history{Received December 3, 2018, accepted February 5, 2019, date of publication February 28, 2019, date of current version March 13, 2019. }

\doi{10.1109/ACCESS.2019.2902122}

\title{Muscle Activity Analysis using Higher-Order Tensor Decomposition: Application to  Muscle Synergy Extraction}

\author{\uppercase{Ahmed Ebied}\authorrefmark{1},  \IEEEmembership{student member, IEEE},
        \uppercase{Eli Kinney-Lang }\authorrefmark{1}, \IEEEmembership{student member, IEEE},
        \uppercase{Loukianos Spyrou} \authorrefmark{1},
                 and \uppercase{Javier Escudero} \authorrefmark{1},\IEEEmembership{Member, IEEE}}

\address[1]{School of Engineering, Institute for Digital Communications, University of Edinburgh, Edinburgh EH9 3FB, United Kingdom}


\markboth
{Ebied \headeretal: Preparation of Papers for IEEE TRANSACTIONS and JOURNALS}
{Ebied \headeretal: Preparation of Papers for IEEE TRANSACTIONS and JOURNALS}

\corresp{Corresponding author: Ahmed Ebied (e-mail: ahmed.ebied@ed.ac.uk).}

\begin{acronym}
\acro{ctd}[consTD]{constrained Tucker decomposition}
\acro{corc}[CORCONDIA]{Core Consistency Diagnostic}
\acro{dof}[DoF]{Degree of Freedom}
\acro{EMG}[EMG] {electromyography}
\acro{NMF}[NMF]{non-negative matrix factorisation}
\acro{ALS}[ALS]{Alternating least square}
\acro{para}[PARAFAC]{Parallel Factor Analysis}
\end{acronym}

\begin{abstract}
Higher-order tensor decompositions have hardly been used in muscle activity analysis despite  multichannel \ac{EMG} datasets naturally occurring as multi-way structures. Here, we seek to demonstrate and discuss the potential of tensor decompositions as a framework to estimate muscle synergies from $3^{rd}$-order \ac{EMG} tensors built by stacking repetitions of multi-channel \ac{EMG} for several tasks.
We compare the two most widespread tensor decomposition models -- \ac{para} and Tucker -- in muscle synergy analysis of the wrist's three main \acp{dof} using the public first Ninapro database. Furthermore, we proposed a \ac{ctd} method for efficient synergy extraction building on the power of tensor decompositions. This method is proposed as a direct novel approach for shared and task-specific synergy estimation from two biomechanically related tasks. Our approach is compared with the current standard approach of repetitively applying \ac{NMF} to a series of the movements.
The results show that the \ac{ctd} method is suitable for synergy extraction compared to \ac{para} and Tucker. Moreover, exploiting the multi-way structure of muscle activity, the proposed methods successfully identified shared and task-specific synergies for all three \acp{dof} tensors. These were found to be robust to disarrangement with regard to task-repetition information, unlike the commonly used \ac{NMF}.
In summary, we demonstrate how to use tensors to characterise muscle activity and develop a new \ac{ctd} method for muscle synergy extraction that could be used for shared and task-specific synergies identification. We expect that this study will pave the way for the development of novel muscle activity analysis methods based on higher-order techniques.
\end{abstract}


\titlepgskip=-15pt

\maketitle
\acresetall 

\section{Introduction}

\PARstart{H}{igher-order} tensors are the generalisation of  vectors ($1^{st}$-order tensors) and matrices ($2^{nd}$-order tensors). When analysing and extracting patterns from higher-order data (data indexed by more than two variables), tensor decompositions may provide several advantages, such as compactness, uniqueness of decomposition, and generality of the identified components, over classical matrix (i.e., $2^{nd}$-order) factorisations \cite{Cichocki2014}. Tucker \cite{Tucker1966a} and \ac{para} \cite{Harshman1970a} are the most widespread methods to factorise the tensor into its main components.

Recently, higher-order tensor decompositions have received substantial attention in biomedical signal processing applications. For instance, they have been utilised frequently in brain activity analysis \cite{Cong2015}. Some applications include analysing electroencephalogram data to classify epileptic patients \cite{Spyrou2016} and analysis of magnetoencephalogram  activity in Alzheimer's disease \cite{Escudero2015}.
Surprisingly, tensor factorisation had hardly been used in \ac{EMG} analysis \cite{Ebied2017}. Recently, we classified wrist movements using the components of a $4^{th}$-order muscle activity tensor to provide a proof-of-concept for the use higher-order tensor decomposition in muscle synergy analysis \cite{Ebied2017}. Another study used Tucker decomposition for feature extraction from a 2-channel  \ac{EMG} for classification \cite{Xie2013a}. Moreover, Delis \textit{el al.}~\cite{Delis2014,Delis2015,Hilt2018} proposed  a space-by-time decomposition model to extract concurrent spatial and temporal components from single-trail \ac{EMG} recordings using a Sample-Based Non-negative Matrix Trifactorization algorithm that resembles a Tucker2 tensor decomposition model \cite{Tucker1966a}. However, a detailed evaluation of the potential of different tensor factorisation models for \ac{EMG} analysis is lacking.

Multichannel \ac{EMG} data are most often represented in matrix form with $time$ and $channels$ as indices along each mode (dimension) so that two-way signal processing methods (i.e., matrix factorisations) are used for muscle activity analysis. However, in most \ac{EMG} studies, data are naturally structured with more modes than the $temporal (samples)$ and $spatial (channels)$ indices. For instance, the \ac{EMG} datasets usually includes repetitions of subjects and/or movements. This means that the muscle activity naturally fits into a higher-order tensor model including additional modes to the $temporal$ and $spatial$ ones. This is what the  studies \cite{Xie2013a,Ebied2017} illustrated and shows that current $2^{nd}$-order approaches do not take advantage of the natural data structure. This means that some information about the interaction between modes may be lost in those approaches. Thus, we hypothesise that higher-order tensor decomposition will be beneficial for muscle activity analysis.

Therefore, our main aim is to explore the use of higher-order tensor decomposition in muscle activity analysis. We propose a \ac{ctd} model for muscle synergy analysis and compare it with the most prominent tensor decomposition models (\ac{para} and Tucker). Hence, we formulate an appropriate and efficient approach for consistent and meaningful muscle synergy extraction. Our secondary objective is to demonstrate the possible application and benefits of tensor decomposition in muscle synergy analysis. Thus, the \ac{ctd} model is utilised to identify shared and task-specific muscle synergies as an illustration for advantages of higher-order tensor decomposition in muscle synergy extraction.

\subsection{Muscle synergy extraction}

\begin{figure}[b]
	\centering
	\includegraphics[width=\linewidth]{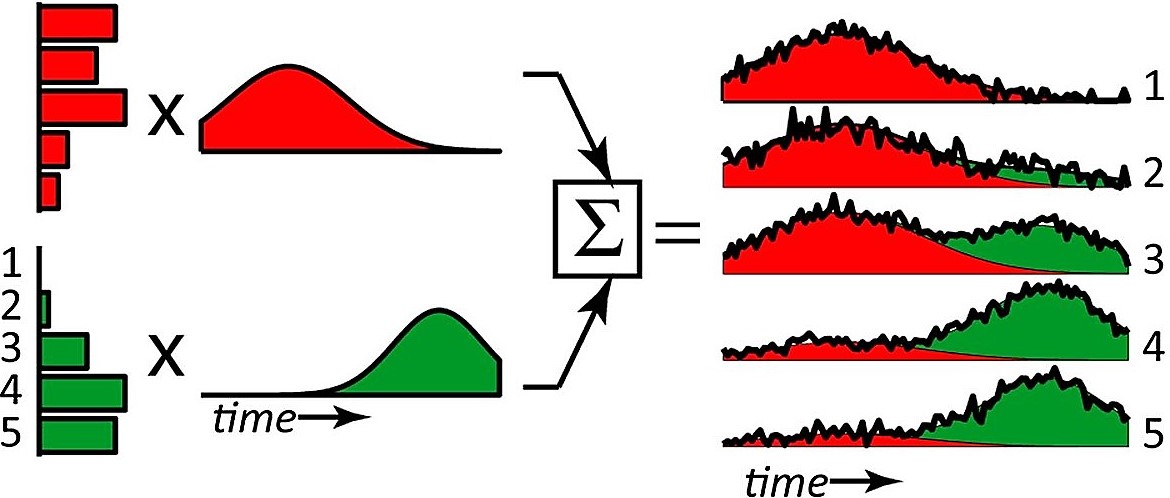}
	\caption{A schematic illustration of muscle synergies. Two muscle synergies (red and green) and their weighting function generate 5-channel \ac{EMG} signal as linear combination. The two colours (red and green) in the \ac{EMG} recording shows how each synergy contributes to the waveform (black line) of each channel. Figure from \cite{Cheung2012b}.}
	\label{fig:SynergyExample}
\end{figure}

The muscle synergy concept  \cite{Tresch1999,Saltiel2001,DAvella2003} provides an explanation for how the central nervous system (CNS) deals with the complexity and high dimensionality of motor control for the musculoskeletal system across multiple \acp{dof} \cite{DAvella2015}. The concept posits that the CNS reduces the motor tasks into a lower-dimensional subspace in a modular form. Simply put, the nervous system activates muscles in groups (synergies) for motor control rather than activating each muscle individually. Hence, the multichannel EMG signal is considered as a linear mixture of muscle synergies with weighting function or activation coefficients across time as illustrated in Figure~\ref{fig:SynergyExample}.

Despite the debate about the neural origin of muscle synergies \cite{Kutch2012,Bizzi2013,TRESCH}, they have  proved useful for many applications such as clinical research \cite{Pons2016a}, prosthesis control \cite{Rasool2016,Ison2015}, and biomechanical studies \cite{Nazifi2017,Martino2015}.

The time-invariant mathematical modelling of muscle synergies \cite{Tresch1999,Saltiel2001} expresses the multi-channel \ac{EMG}  as a combination of synchronous synergies scaled by a set of respective weighting functions. This leads to the formulation of a blind source separation problem represented as a matrix factorisation where several techniques have been explored, including \ac{NMF} \cite{Lee1999},  PCA \cite{Jackson1991} and ICA \cite{Hyvarinen2000}. Among them, \ac{NMF} is the most prominent and suitable method  \cite{Ebied2017,Tresch2006} since the non-negativity constraint makes it more appropriate and easily interpretable due to the additive nature of synergies \cite{Choi2011}. However, all these approaches  are 2$^{nd}$-order analysis methods. This may limit them when dealing with situations where repetitive analysis are investigated such as the identification of shared muscle synergies.

\subsection{Shared  synergies identification}

The notion of shared synergies derives directly from the muscle synergy concept. This implies that shared synergies can be found in diverse motor tasks sharing some mechanical or physical characteristics. Support for this idea comes from animal studies (frogs\cite{DAvella2005,Cheung2005} and cats \cite{Torres-Oviedo2006}) as well as human studies where shared and task-specific synergies distinctive for one motor task or movement have been investigated across activities such as walking and cycling \cite{Barroso2014}, postural balance positions \cite{Chvatal2011a,Torres-Oviedo2010}, and normal walking and slipping \cite{Nazifi2017,Martino2015}.

The current approach to estimate the shared and task-specific synergies is to apply \ac{NMF} on the multi-channel \ac{EMG} signals recorded during the tasks in question. This is done for several repetitions of each task and usually for a number of different subjects. Then, the synergies are rearranged across tasks, repetitions and subjects (in some cases) in order to maximise the similarity between a set of synergies,  which is assumed to be shared across tasks and/or subjects. Most of the shared or common synergies identification studies rely only on correlation coefficients as a similarity metric to differentiate between shared and task-specific synergies. Nonetheless, this approach is limited by the fact that the rearrangement of synergies would have a significant effect on identifying shared synergies \cite{Barroso2014,Martino2015,Nazifi2017}. 

Thus, the second objective of this manuscript is to devise the \ac{ctd} method to take advantage of the multi-way structure in \ac{EMG} activity to extract shared muscle synergies. Strictly speaking, the concept of muscle synergy is applicable to 2$^{nd}$-order data but we are here inspired by it to extract analogous synergies via tensor factorisation. The \ac{ctd} method will be compared against the current traditional method that uses $2^{nd}$-order analysis model (\ac{NMF}). We illustrate this with wrist movements.

\section{Materials}

In this study, we analyse surface \ac{EMG} datasets from the publicly-available Ninapro first database \cite{Atzori2014,Atzori2015a}. The data were collected from 27 healthy subjects instructed to perform 53 wrist, hand and finger movements with 10 repetitions for each movement. The ``stimulus" time series in the Ninapro dataset is used to set the segment's start and end points for each movement repetition. Each segment consists of 10-channel surface \ac{EMG} signals recorded by  a MyoBock 13E200-50 system (Otto Bock HealthCare GmbH), rectified by root mean square and sampled at 100Hz.

A selected number of wrist movements are included in this study since we will use upper-limb myoelectric control as an impactful vehicle to demonstrate the techniques. This application of muscle synergy has received substantial attention recently  \cite{Choi2011,Jiang2014b,Ma2015a,Lin2017} and we expect that its selection to illustrate our study will promote uptake of the methods. Biomechanically related tasks/movements were chosen to identify shared and task-specific synergies between them. Related tasks for this study focused specifically on wrist motion and its three main \acp{dof}. Concerning to the wrist movements, 3 \acp{dof} are always considered in  myoelectric control radial and ulnar deviation (DoF1) , wrist extension and flexion (DoF2) and finally wrist supination and pronation (DoF3). The three \acp{dof} represent the horizontal, vertical and rotation \acp{dof} or movements   respectively.

\section{Methods}

\subsection{Higher-order tensor decomposition}

\subsubsection{Tensor construction}

Tensors are a higher-order generalisation of vectors (1$^{st}$-order) and matrices (2$^{nd}$-order). The first step to create a higher-order synergy model is to prepare the data in higher order tensor form. This process of  transformation or mapping lower-order data to higher-order data is known as ``tensorisation". Several  stochastic and deterministic techniques have been used for tensorisation \cite{Debals2015}. ``Segmentation" is one of the deterministic techniques where lower-order tensors are reshaped into higher-order form by segmenting the data into smaller segments and stacking them after each other.

\begin{figure*}[t!] 
	\begin{subfigure}{0.4\textwidth}
		\includegraphics[width=\linewidth]{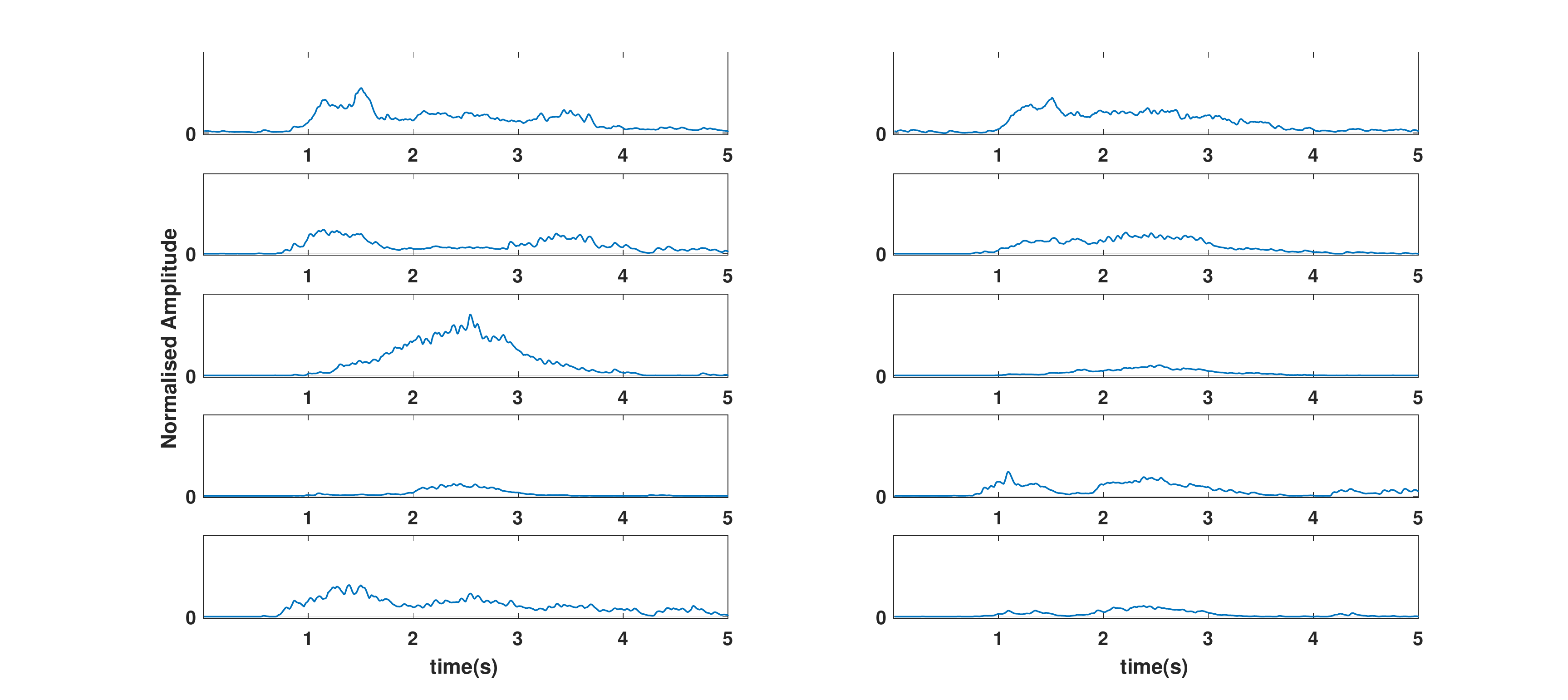}
		\caption{\textit{1-repetition of ulnar deviation}} \label{fig:exampleRep1task16}
	\end{subfigure}\hspace*{\fill}
	\begin{subfigure}{0.4\textwidth}
		\includegraphics[width=\linewidth]{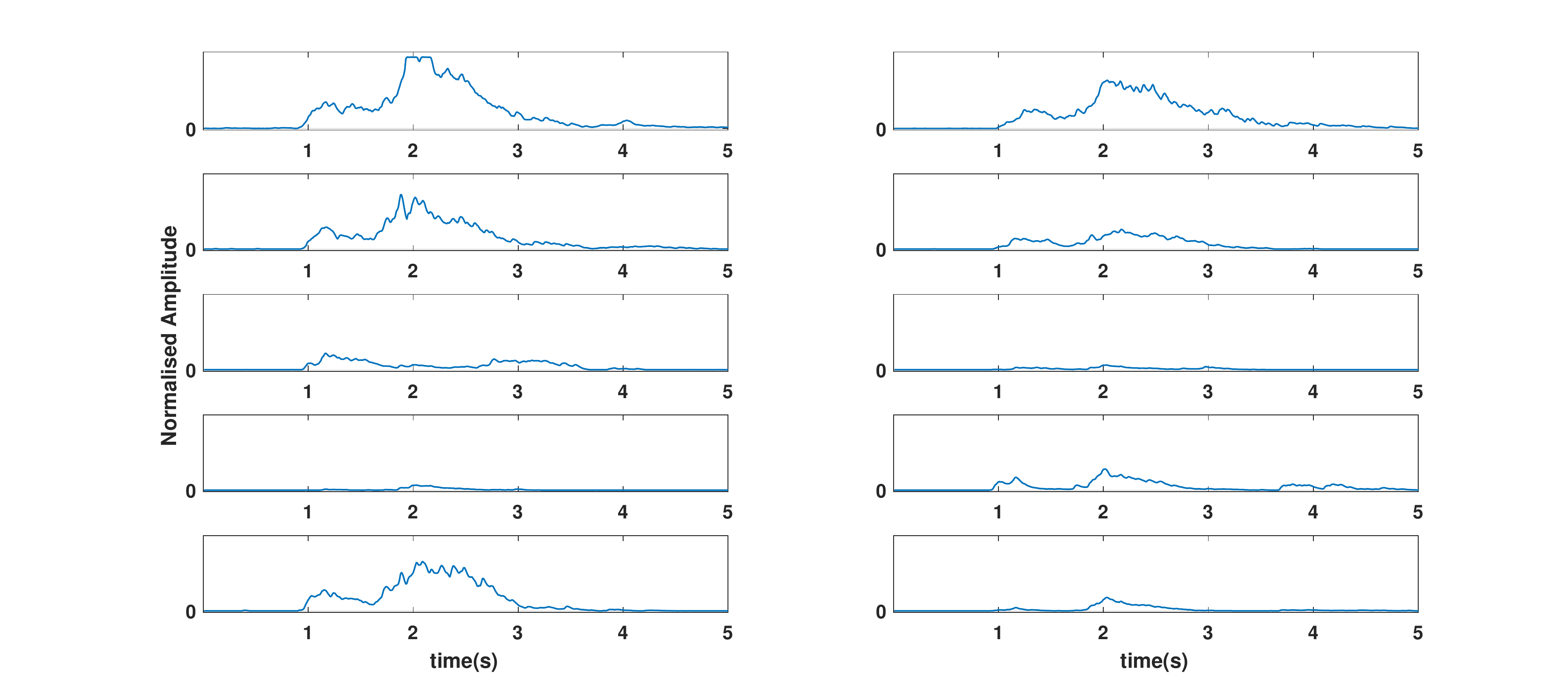}
		\caption{\textit{1-repetition of radial deviation}} \label{fig:exampleRep1task17}
	\end{subfigure}
	
	\medskip
	\begin{subfigure}{0.4\textwidth}
		\includegraphics[width=\linewidth]{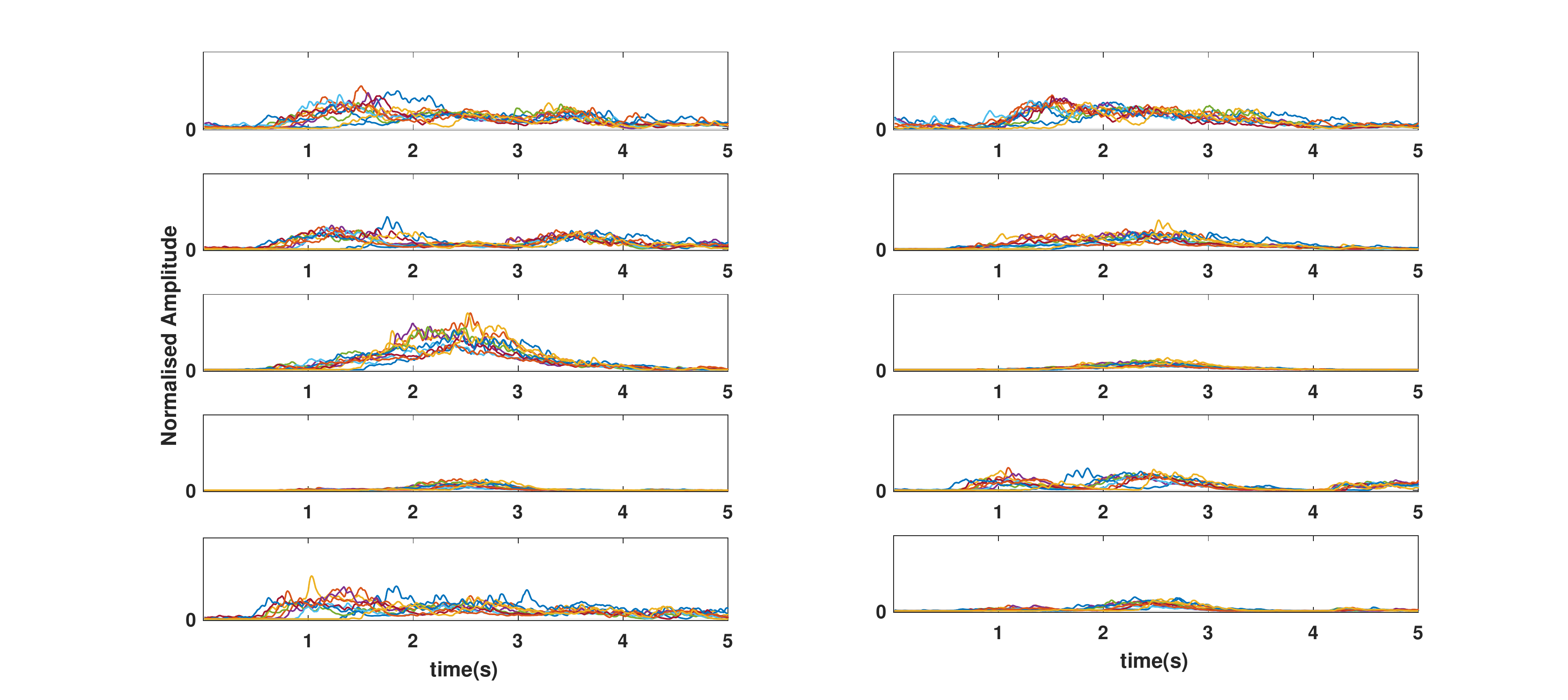}
		\caption{\textit{10-repetition of ulnar deviation}} \label{fig:exampleRep10task16}
	\end{subfigure}\hspace*{\fill}
	\begin{subfigure}{0.4\textwidth}
		\includegraphics[width=\linewidth]{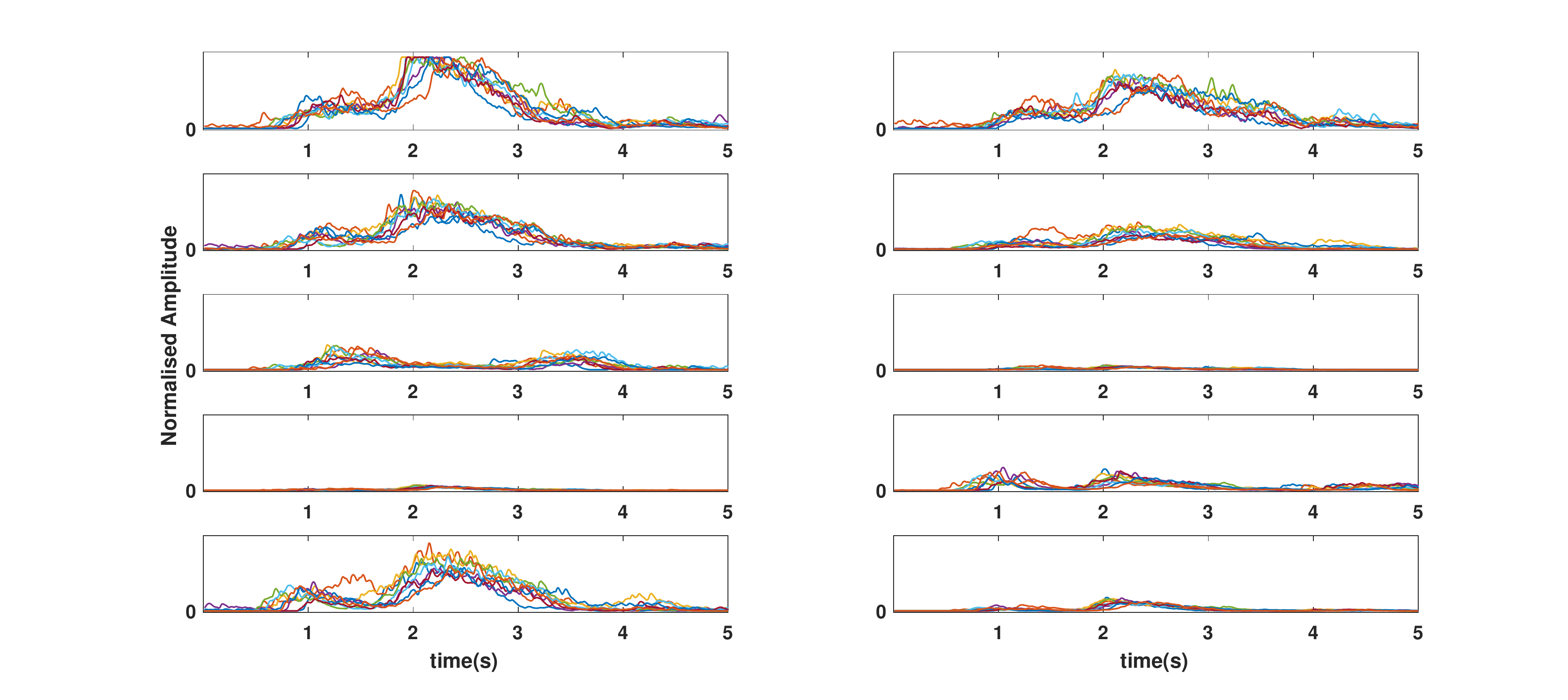}
		\caption{\textit{10-repetition of radial deviation}} \label{fig:exampleRep10task17}
	\end{subfigure}

			\centering
	\begin{subfigure}{\textwidth}
		\centering
		\includegraphics[width=0.4\linewidth]{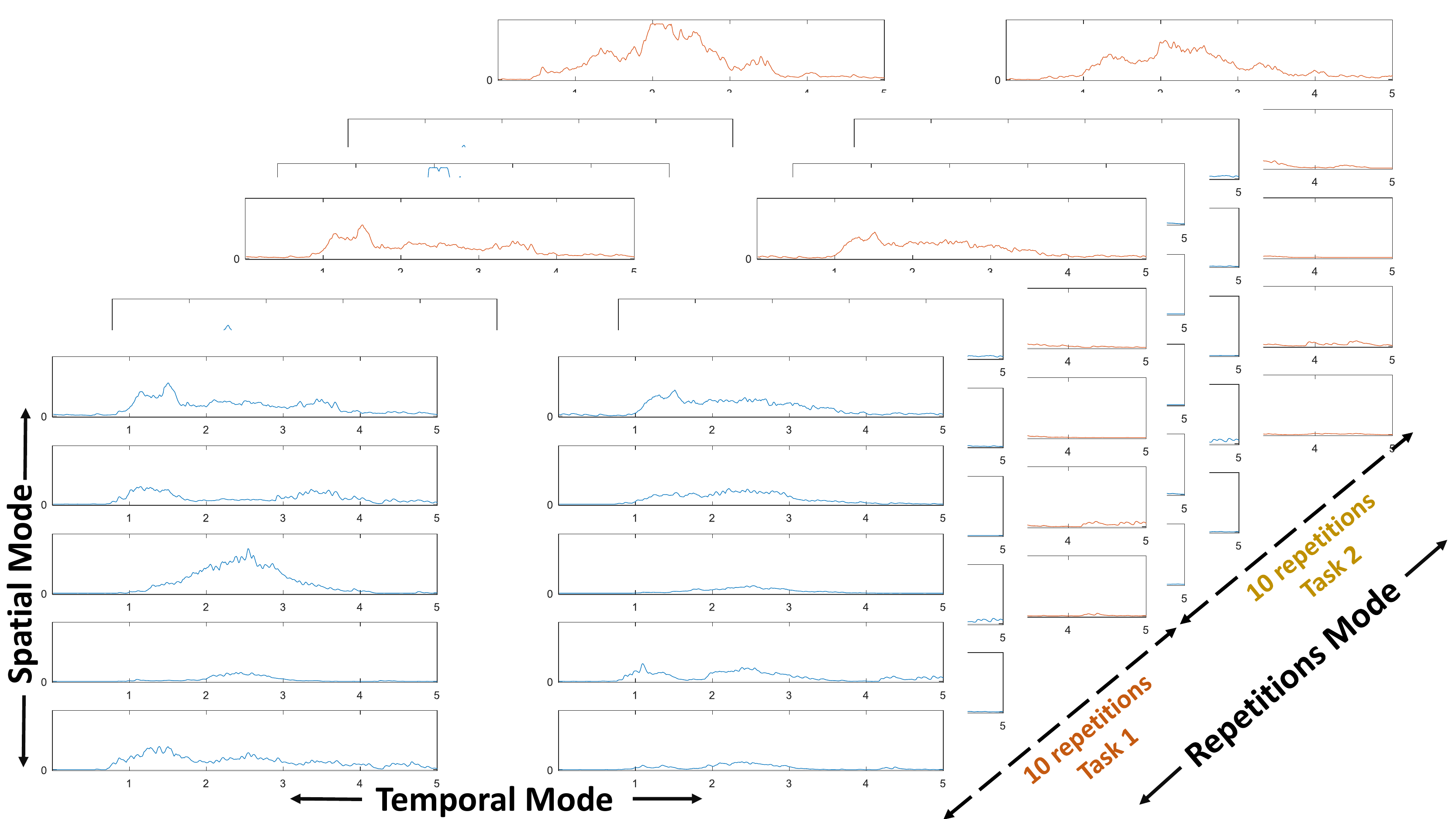}
		\captionsetup{justification=centering,margin=3cm}
		\caption{ \textit{$3^{rd}$-order tensor for radial and ulnar deviations movements with modes [ 500-samples (5-seconds) $\times$ 10-channels$\times$ 20-repetitions (10 repetitions for each movements)}]} \label{fig:exampleTensor}
	\end{subfigure}\hspace*{\fill}
	
	\caption{An example for data construction and tensor decomposition. Figures \ref{fig:exampleRep1task16} and \ref{fig:exampleRep1task17} are  10-channel EMG recordings of Ninapro first database (subject ``1"). The recording is for one repetition of ulnar (\ref{fig:exampleRep1task16}) and radial deviation (\ref{fig:exampleRep1task17}) movements (DoF1). (Figures \ref{fig:exampleRep10task16} and \ref{fig:exampleRep10task17}) shows 10 repetitions of each movements, which are stacked together to form a $3^{rd}$-order tensor as in \ref{fig:exampleTensor} for DoF1. } 
	\label{fig:ExampleContructionTensor}
\end{figure*}

To create  3$^{rd}$-order tensor  for \ac{EMG} dataset,  the multichannel \ac{EMG} recordings of several movements and/or tasks can be represented as a matrix with \textit{time} and \textit{channels} are its dimensions or modes.  This matrix are segmented into equal epochs where each epoch contains one repetition of one movement or task. By stacking these epochs across the new \textit{repetition}  mode, a 3$^{rd}$-order tensor is created with modes  $temporal\times spatial\times repetition$.  A given wrist's \ac{dof} tensor is constructed by stacking repetitions of wrist movements. For example, a 1-\ac{dof} tensor is created using ulnar (Figure \ref{fig:exampleRep10task16}) and radial (Figure \ref{fig:exampleRep10task17}) deviation movements repetitions to form a $3^{rd}$-order tensor as shown in Figure \ref{fig:ExampleContructionTensor}. The 2-\acp{dof} tensor consists of 4 wrist movements repetitions; the ulnar  and radial deviation in addition to wrist extension and flexion movements. Both tensors are used in the comparison between tensor decomposition models. However, only the 1-\ac{dof} tensors  are used for shared and task-specific synergies comparison against \ac{NMF} for simplicity as we introduce this application as a proof of concept.

\subsubsection{Tensor decomposition models}

\begin{figure}[t]
	\centering	
	\begin{subfigure}[b]{0.49\textwidth}
		\centering
		\includegraphics[width=\linewidth]{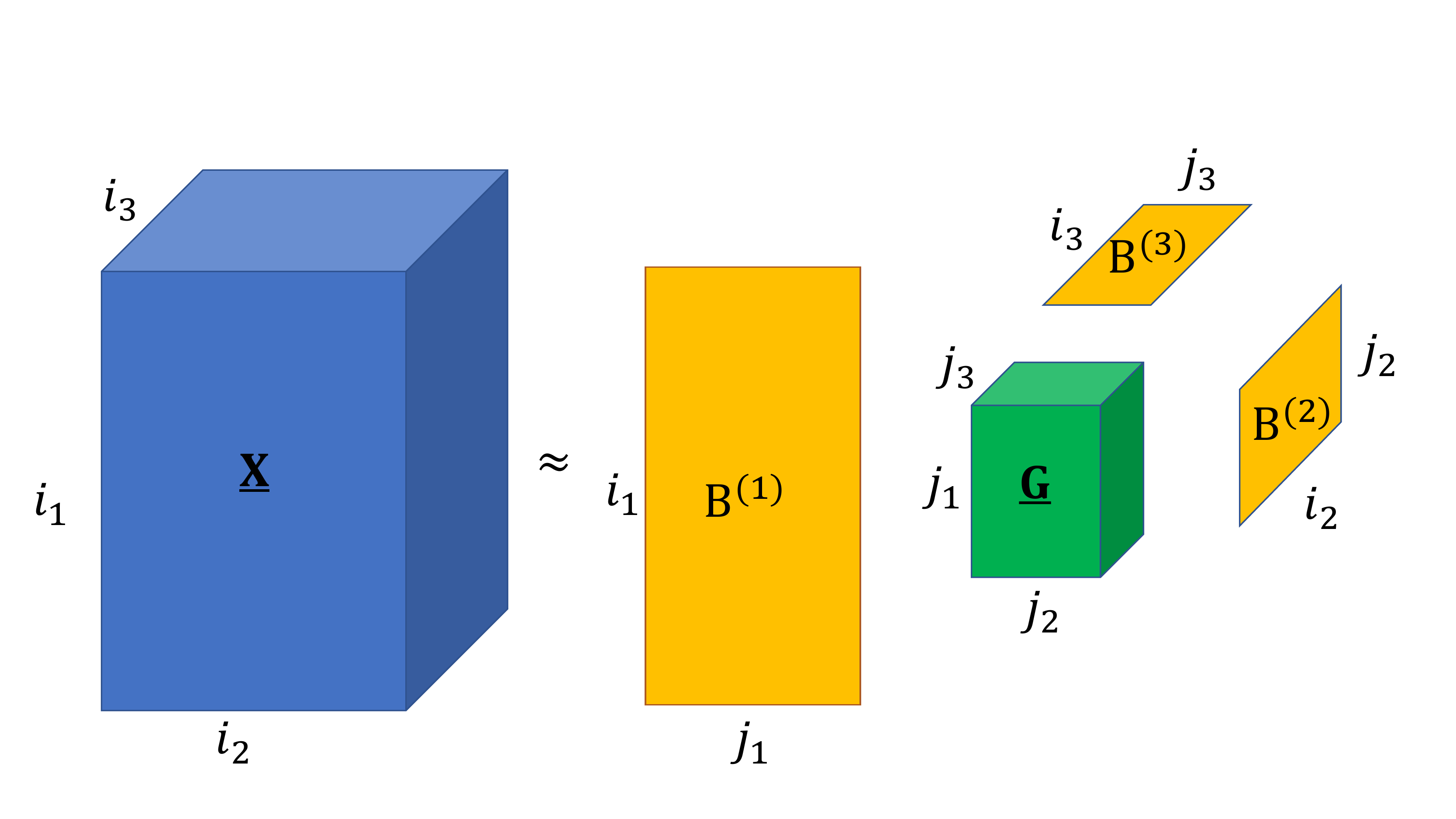}
		\caption{}
		\label{fig:Tuckergraphics}
	\end{subfigure}
	\hfill
	\begin{subfigure}[b]{0.49\textwidth}
		\centering
		\includegraphics[width=\linewidth]{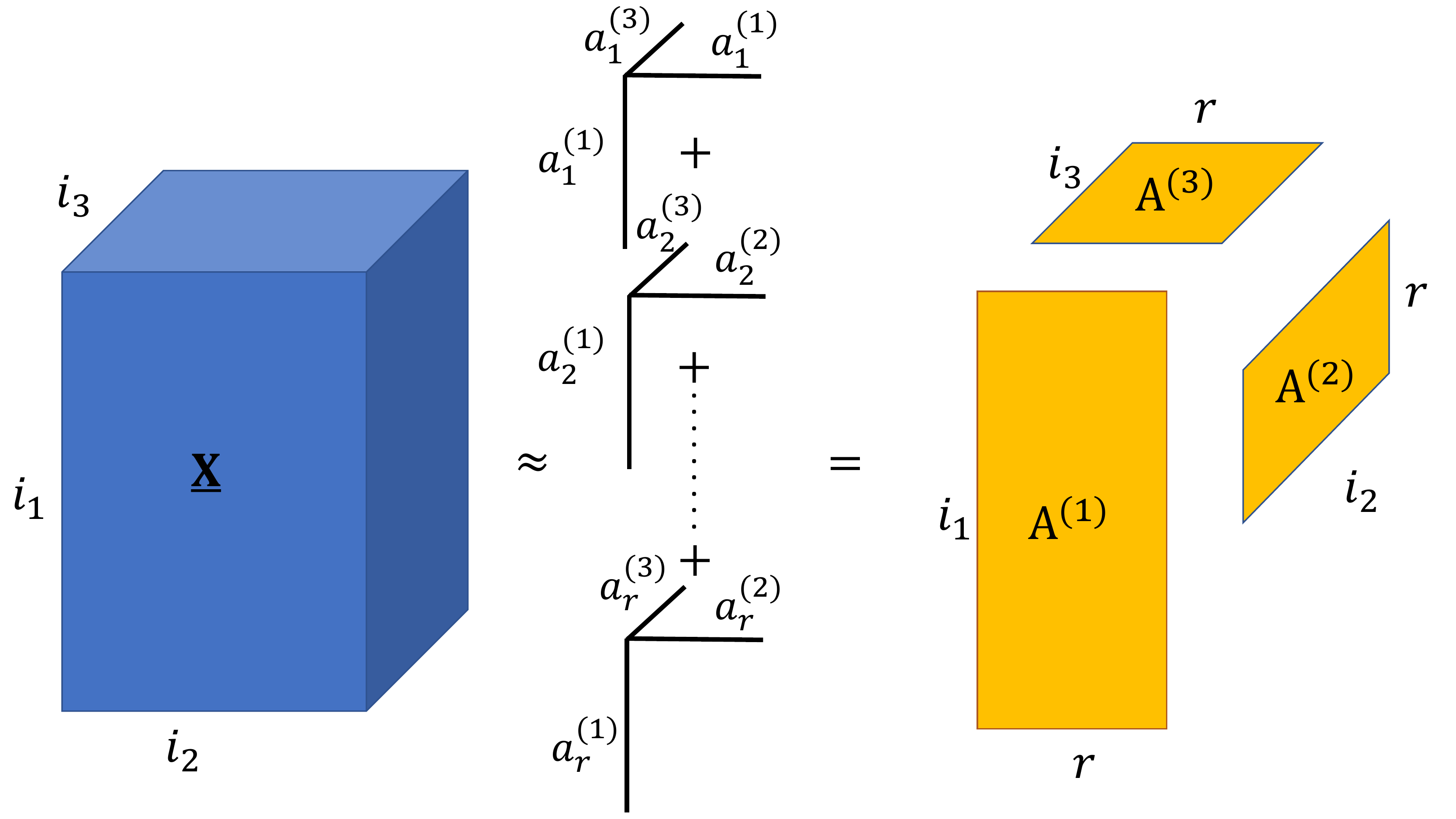}
		\caption{}
		\label{fig:parafacgraphics}
	\end{subfigure}
	\caption{Illustration of  Tucker (\ref{fig:Tuckergraphics})  and  PARAFAC (\ref{fig:parafacgraphics}) decomposition for $3^{rd}$-order tensor $\underline{\mathbf{X}}$.}
	\label{fig:ParafacAndTuckerGraphics}
\end{figure}

Higher-order tensors can be decomposed into their main components (also known as factors) in a similar way to matrix factorisation methods. Several tensor decomposition models have been introduced with Tucker and \ac{para} being the most prominent ones \cite{Comon2014}.  In Tucker decomposition, the higher-order tensor is decomposed into a smaller core tensor transformed by a matrix across each mode, where the core tensor determines the interaction between those matrices. On the other hand, \ac{para} could be considered as a restricted case of Tucker with a supra-diagonal core tensor with 1s across its supra-diagonal and 0s elsewhere, which is also known as an ``identity tensor". Consequently, the number of components is fixed for all modes for \ac{para}, unlike Tucker where the numbers of components in the different modes, can differ.  Thus, the Tucker model has more flexibility in component number unlike \ac{para}  \cite{Kolda2008c}. The differences between both models are represented in Figure~\ref{fig:ParafacAndTuckerGraphics}.

In general, an $n^{th}$-order tensor $\underline{\mathbf{X}}\in \mathbb{R}^{i_{1}\times i_{2} \times .... i_{n}}$ can be factorised according to the Tucker model as follows:
\begin{equation}\label{eq_tucker_model}
\underline{\mathbf{X}} \approx \underline{\mathbf{G}} \times_{1}\mathbf{B}^{(1)} \times_{2}\mathbf{B}^{(2)} \dots \times_{n}\mathbf{B}^{(n)} 
\end{equation}
where $\underline{\mathbf{G}}\in \mathbb{R}^{j_{1}\times j_{2} \dots \times j_{n}} $ is the core tensor and   $\mathbf{B}^{(n)}\in \mathbb{R}^{i_{n}\times j_{n}}$ are the component matrices transformed across each mode while ``$\times_{n}$" is  multiplication  across the $n^{th}$-mode \cite{Kolda2008c}. The core tensor $\underline{\mathbf{G}}$ is flexible to have different dimensions across each mode as long as it is smaller than the tensor being decomposed, $\underline{\mathbf{X}}$, so that $j_{n} \le i_{n}$. 

On the other hand, the \ac{para} approach factorises the $n^{th}$-order tensor $\underline{\mathbf{X}}\in \mathbb{R}^{i_{1}\times i_{2} \times \dots i_{n}}$ into its component matrices with fixed number of components across each mode as
\begin{equation}\label{eq_PARAFAC_model}
\underline{\mathbf{X}} \approx \underline{\mathbf{\Lambda}} \times_{1}\mathbf{A}^{(1)} \times_{2}\mathbf{A}^{(2)} \dots \times_{n}\mathbf{A}^{(n)} 
\end{equation}
where $\underline{\mathbf{\Lambda}}\in \mathbb{R}^{r\times r \dots\times r_{n}}$ is a super diagonal tensor that have same dimension across each mode and the vector \bm{$\lambda$}  is across the diagonal of $\underline{\mathbf{\Lambda}}$. This limits the interactions in-between components unlike Tucker decomposition. The \ac{corc} is an index to assess the appropriateness of \ac{para} decomposition by measuring the the degree of super-diagonally \cite{Bro2003}. The core consistency is less than or equal to 100\% where consistency close to 100\% implies an appropriate multilinear model, whereas a lower core consistency (lower than 50\%) would mean a problematic or even invalid model \cite{Bro2003}.

The \ac{para} decomposition is unique under very mild conditions \cite{Kolda2008c}. On the other hand, the Tucker decomposition generally does not provide unique solutions \cite{Cichocki2014}. However, the uniqueness for the Tucker model can be achieved in practice by imposing additional constraints on the modes \cite{Zhou2012}.

\subsubsection{Alternating Least Squares algorithm}
\label{sec:ALS}

Both Tucker and \ac{para} can be estimated with the \ac{ALS}. \ac{ALS} starts initialising the components (and core tensor in the case of Tucker decomposition) to be estimated either randomly  \cite{Harshman1994} or by using other methods such as singular value decomposition or direct trilinear decomposition \cite{Sands1980}. After initialisation, the next step is the iteration  phase to minimise the loss function between the original data and its model by breaking down this complex non-convex problem into a series of simpler, convex problems,  which are tackled in succession \cite{Comon2009}. This is done by fixing all the component matrices to be estimated except for those corresponding to one of the modes and alternate iteratively between all the components to solve each convex problem until convergence \cite{Cichocki2009b}.

In the case of the $3^{rd}$-order \ac{EMG} tensor $\underline{\mathbf{X}}\in \mathbb{R}^{i_{1}\times i_{2} \times i_{3}}$ , The Tucker model equation \ref{eq_tucker_model} would be expressed as 
\begin{equation}\label{eq_tucker_model_ALS_example}
\underline{\mathbf{X}} \approx \underline{\mathbf{G}} \times {} _{1}\mathbf{B}^{(1)} \times {} _{2}\mathbf{B}^{(2)} \times {} _{3}\mathbf{B}^{(3)} 
\end{equation} 
where $\mathbf{B}^{(1)} \in \mathbb{R}^{i_{1}\times j_{1}}$ is the \textit{temporal} mode while $\mathbf{B}^{(2)} \in \mathbb{R}^{i_{2}\times j_{2}}$ and $\mathbf{B}^{(3)} \in \mathbb{R}^{i_{3}\times j_{3}}$ are the \textit{spatial} and \textit{repetition} modes respectively. $j_{1},j_{2}$ and $j_{3}$ are the number of components in each mode and  the core tensor $\underline{\mathbf{G}}$ dimensions as well. The least squares loss function for this model would be
\begin{equation}\label{eq_loss_fun_Tucker}
 argmin_{\mathbf{B}^{(1)},\mathbf{B}^{(2)},\mathbf{B}^{(3)},\mathbf{\underline{G}}} \|\mathbf{X^{(i_{1}\times i_{2}i_{3})}-\mathbf{B}^{(1)}\underline{\mathbf{G}}(\mathbf{B}^{(3)}\otimes \mathbf{B}^{(2)})^{T}} \|^{2}
\end{equation}
 where $\otimes$ is a Kronecker product, a generalisation of the outer product, that can be applied on matrices of arbitrary size \cite{Liu2008}. The algorithms in this study are based on adapted \ac{para} and Tucker functions from the (N-WAY Toolbox) for Matlab \cite{Andersson2000}. 

The \ac{ALS} algorithm has several advantages such as simplicity compared to the simultaneous approaches. However, it is not guaranteed to converge to a stationary point as the problem could have several local minima. Therefore, multiple constraints on initialisation and iteration phases would help to improve the estimation \cite{Kolda2008c}. Moreover, constraining the tensor models has several benefits including: improving the uniqueness of the solution, more interpretable results that do not contradict \textit{a priori} knowledge, avoiding degeneracy and numerical problems, and speeding up the algorithm. Although constraints may lead to poorer fit for the data compared to the unconstrained model, the advantages outweigh the decrease in the fit for most cases \cite{Cichocki2014}. The decomposition models are constrained through their \ac{ALS} algorithm in the initialisation and/or iteration phases. For example, non-negativity constraint is one of the most commonly used ones due to the illogical meaning for negative components in many cases. The non-negativity constraint is implemented in the updating step by setting the negative values of computed components to zero by the end of each iteration to force the algorithm to converge into a non-negative solution. In the case of muscle synergy extraction, non-negativity would add more information to the decomposition by taking in account the additive nature of muscle synergies.

\subsection{Constrained Tucker decomposition}
\label{sec:const_tuck}

In this section, the \ac{ctd} method is discussed in detail, including number of components and different constraints imposed to improve muscle synergy extraction.

Imposing constraints on the tensor decomposition model has several benefits  \cite{Cichocki2014} as discussed in Section~\ref{sec:ALS}.
Therefore, in order to extract consistent and meaningful muscle synergies, a \ac{ctd} model is proposed for muscle synergy analysis. We hypothesise  that this model would benefit from the flexibility and versatility of Tucker model in comparison with \ac{para} decomposition while retaining the high explained variance. In addition, the additional constraints would result a unique and consistent synergy extraction.  This approach was inspired by the shared-synergy concept \cite{DAvella2005,Cheung2005} by including  additional components to account for any shared variability across movements, tasks or \acp{dof}.

    \subsubsection{Number of components} 
    \label{sec:const_tuck_numComp} 
In this setup, one $temporal$ component is assigned for each \ac{dof} instead of two components as preliminary results showed that when assigning 2 components for one DoF, the temporal activity would be segmented as the following: one component will capture the main activity (middle of the segment) will the other will capture the rest period (at the beginning and end of the segments). Therefore, the additional temporal component will not be beneficial  for synergy extraction since we are concerned with extracting the main muscle activity  to identify shared synergy for each \ac{dof}. In addition, we are aiming to reduce the number of elements in the decomposition focusing on the main activity. 

The number of $spatial$ mode components (synergies) were chosen according to the functional approach (discussed in Section \ref{sec:number_syn}). In general, we aim for a parsimonious model that could extract meaningful muscle synergies with the least number of components and elements. Hence, two components are assigned to each \ac{dof}  to estimate a task-specific synergy for each movement  for both $spatial$ and $repetition$ modes. In addition, an additional component (shared) was assigned in these two modes in order to improve the data fit and account for any common variability inspired by the shared synergy concept. 

    \subsubsection{Additional constraints}

 Four constraints on the Tucker model were proposed to facilitate the muscle synergy identification. Two constraints are imposed in the initialisation phase on  the core tensor and  $repetition$ mode components, while the other two are applied during the iteration phase of \ac{ALS} algorithm. 

The core tensor is initialised to link each of the components in the $temporal$ and $repetition$ modes into their respective $spatial$ components (synergy). The core tensor is initialised and fixed (does not update with each iteration) into a value of $1$ between each spatial synergy and its respective components in the other modes and $0$ otherwise. This ensures that every spatial synergy is assigned to only one $repetition$ component and avoids any cross interaction. The values of the core tensor are chosen to be $1$ to account for  all the variability in the mode components. This fixed sparse core tensor setup controls the interactions between components in each mode and links the repetition components to its respective components it $temporal$ and $spatial$ modes.  

In addition, since we know that each repetition in the 3$^{rd}$-order tensor belongs to a known movement, we use this information in tensor decomposition by constraining the  $repetition$ mode. The components of  the $repetition$ mode are initialised and divided into ``task-specific" and ``shared" components. The task-specific  components are initialised to $1$ for a repetition of the considered movement or task and $0$ otherwise, while the shared component is initialised by a  value of $0.5$ for all repetitions. Unlike the core tensor the update of this mode is not fixed to account for the variability and differences between repetitions of the same movement, alternatively, a controlled averaging constraint is used during the iteration  phase. The controlled averaging and repetition mode initialisation works together to incorporate the repetitions information into the tensor decomposition and help it to identify the shared and task-specific synergies separately.

The other two constraints on updating components in Tucker's \ac{ALS} algorithm are the non-negativity on $temporal$ and $spatial$ modes and the controlled averaging on the initialised $repetition$ mode. The non-negativity constraint is imposed in order to have meaningful components (synergies) \cite{Choi2011,Ebied2017} as discussed in Section~\ref{sec:ALS}. 

The controlled averaging constraint aims to allow some variability within each $repetition$ component whether it is shared or task-specific.  This approach will hold the structure of $repetition$ factors that was initialised without fixing it through iterations and take the differences between repetitions into account. As for controlled averaging implementation, it is a simple moving-averaging filter with window length ($k=3$) implemented by modifying the iteration phase in the \ac{ALS} algorithm. It is applied on \textit{repetition} mode components at the end of each iteration to account for the variability between each repetition, hence, increase the explained variance.   We acknowledge this is a simple implementation but we expect it to be representative of our procedure while achieving a good performance in muscle synergy identification.

\subsection{Tensor decomposition  models for muscle synergy analysis}

\subsubsection{Number of Synergies}
\label{sec:number_syn}

Selecting the appropriate number of components (including synergies in  $spatial$ mode) for higher-order tensor models is instrumental in capturing the underlying structure of the data \cite{Cichocki2009b}. Several mathematical approaches have been deployed to determine the appropriate number of components for higher-order tensor decomposition such as \ac{corc} \cite{Bro2003}, heuristic and approximating \cite{Nie2009} techniques.

On the other hand, the number of synergies for $2^{nd}$-order model extracted via matrix factorisation methods have been determined using two main approaches: a functional approach, and a mathematical approach \cite{Ebied2018}. The functional approach relies on prior knowledge of data structure and myoelectric control requirements to choose the appropriate number of synergies. For instance, one \cite{Muceli2012c} or two \cite{Jiang2009} synergies are assigned to each wrist's \ac{dof} for proportional myoelcetric control. The mathematical approaches are similar to methods traditionally used in higher-order tensor models. It relies mathematical computation such as explained variance  or the likelihood criteria  \cite{Ikeda2000}.

In order to choose the number of components for \ac{para} and Tucker models, the prior knowledge of the data structure (\textit{i.e.,} number of movements) had been utilised as in the functional approach of matrix factorisation. In addition, the  mathematical criteria (\ac{corc} and explained variance) was used to test and compare different number of components. Both the 1-\ac{dof} and 2-\acp{dof} tensors were decomposed using \ac{para} with a set number of components (2, 3 and 4 components) since we aim to estimate at least one synergy for each movement and  the number of movements are 2 and 4 in the 1-\ac{dof} and 2-\acp{dof} tensors respectively. This helped to guarantee each movement was identified by at least one muscle synergy. Similarly, [2,2,2], [3,3,3] and [4,4,4] Tucker models were used to decompose both tensors. In order to compare both models (Tucker and \ac{para}), number of components for the Tucker model was the same in all modes to match the \ac{para} model for comprehensive comparison.

\subsubsection{Tucker and PARAFAC models for synergy extraction} 

In order to examine the use of Tucker decomposition for muscle synergy extraction, both tensors (1-\ac{dof} and 2-\acp{dof}) were decomposed with a non-negative Tucker decomposition. Three models were applied on the 3$^{rd}$-order tensors with different number of components; [2,2,2], [3,3,3] and [4,4,4]. The time  for algorithm execution is recorded for every run  across the 27 subjects as well as the explained variance percentage as a metrics to compare tensor decomposition models. 

On the other hand, to highlight the differences between Tucker and \ac{para} models in muscle activity analysis, a 2-, 3- and 4-component \ac{para} models with non-negativity constraints are applied on the same wrist's tensors.  The execution time for each run as well as \ac{corc} were recorded to compare between tensor decomposition for synergy analysis. The number of components for both methods were chosen according to the criteria discussed in Section~\ref{sec:number_syn}.

\subsubsection{\ac{ctd} for synergy extraction}      

The proposed \ac{ctd} method is applied to  the 1- and 2-\acp{dof} tensors for muscle synergy analysis. The \ac{ctd} aims to extract one synergy for each movement (task-specific synergy) in addition to a shared synergy across all movements. The number of components for \ac{ctd} were  $[1,3,3]$ for 1-\ac{dof} tensors and $[2,5,5]$ for 2-\acp{dof} tensors according to the criteria discussed in Section \ref{sec:const_tuck_numComp}. 

Therefore, two components in the $spatial$ and $repetition$ modes were assigned to each \ac{dof} in addition to an additional ``shared" component in these two modes. While one $temporal$ component is assigned for each \ac{dof} since we do not need to segment main temporal activity. Thus, a $[1,3,3]$ \ac{ctd} is developed to estimate interpretable components from 1-\ac{dof} tensors, while a $[2,5,5]$ model was used for the 2-\acp{dof} tensors. 

\begin{table}[ht]
\centering
\caption{Core tensor intialisation for \ac{ctd} models.}
\label{table_ch5:coreTensor}
\begin{tabular}{cc|cc} 
\toprule
\multicolumn{2}{c|}{$[1,3,3]$} & \multicolumn{2}{c}{$[2,5,5]$} \\ 
\hline
$g_{1,n,n}=1$  & $n\in\left\{1,2,3 \right\}$  & $g_{1,n,n}=1$ & $n\in\left\{1,2,5 \right\}$  \\
 &  & $g_{2,n,n}=1$  & $n\in\left\{3,4,5\right\}$  \\
$g=0$  & $otherwise$  & $g=0$  & $otherwise$  \\
\bottomrule
\end{tabular}
\end{table}

The core tensor is initialised and fixed for both \ac{ctd} models accordingly as shown in Table~\ref{table_ch5:coreTensor}. The $repetition$ mode is initialised as discussed in Section~\ref{sec:const_tuck} where the task-specific components are initialised by $1$ for a repetitions of the considered movement and $0$ otherwise and the shared component is initialised by a  value of $0.5$ for all repetitions. 
The  $repetition$ mode is constrained in the iteration phase through  controlled averaging  while the  non-negativity constraint is imposed on the $temporal$ and $spatial$ modes.

\subsubsection{Experimental settings}

In order to compare the  proposed constrained Tucker model for muscle synergy analysis with non-negative Tucker and \ac{para} models. The three algorithms were run 10 times in order to examine the uniqueness of solution by testing the ability of algorithms to converge to the same point with similar resulting components. For each run, the time of execution and explained variance were recorded for both Tucker decomposition models while \ac{corc} and execution times were recorded for \ac{para}. All  decomposition models are performed using Matlab 9 with Intel core i7 processor (2.4 GHz, 12 GB RAM).

\subsection{Shared synergy identification}

 The \ac{NMF} approach to identify the shared synergies mainly depends on similarity metrics such as correlation coefficient or coefficient of determination $R^{2}$. Synergies are  estimated from repetitions of single tasks either by taking the average \ac{EMG} activity then applying \ac{NMF} \cite{Nazifi2017,Barroso2014}, or by averaging the synergies extracted from each repetition \cite{Torres-Oviedo2010}. The result would be group of synergies for each task. This is  followed by computing correlation coefficients between synergies of different tasks and identifying shared synergies by matching the synergies of highest correlation coefficient.
 
 On the other hand, the tensor approach  stacks the repetitions from  different tasks together to form a $3^{rd}$-order tensor with modes $channels \times time \times repetitions$ as shown in Figure~\ref{fig:ExampleContructionTensor}. Constrained tensor factorisation is applied on this  $3^{rd}$-order tensor to extract the shared synergies as well as the task specific ones without the need of any similarity metric, unlike \ac{NMF} approach.  The \ac{ctd} directly identifies  shared synergy and 2 task-specific synergies in the $spatial$ mode for $3^{rd}$-order tensor with repetitions  of 2 tasks (ulnar and radial deviation, for example). 
 
\subsubsection{NMF as benchmark}

\ac{NMF} \cite{Lee1999} is used in this study as a comparative benchmark for shared synergy identification \cite{DAvella2005,Barroso2014,Nazifi2017,Chvatal2013a}. \ac{NMF} processes the multi-channel \ac{EMG} recording as a matrix $\mathbf{X}$ with dimensions $channel \times time$. \ac{NMF} decomposes \ac{EMG} recordings into two smaller matrices (factors). The first factor holds the temporal information (also known as weighting function) $\mathbf{B}^{(1)}$ while the other is the muscle synergy holding the spatial information $\mathbf{B}^{(2)}$ as 
\begin{equation}
 \mathbf{X} \approx  \mathbf{B}^{(1)}\times\mathbf{B}^{(2)^{T}}
 \end{equation}
where both  $\mathbf{B}^{(1)}$and $\mathbf{B}^{(2)}$ are constrained to be non-negative. For details see  \cite{Devarajan2008}.

Since the dataset had 10 repetitions for each task, \ac{NMF} was applied on each of them. The number of synergies was chosen by variance accounted for (VAF) as a metric \cite{Rasool2016}.  The first step to identify the shared and task-specific synergies would be finding the reference synergy \cite{Nazifi2017,Barroso2014} from the 10 repetitions of that task. This is done by calculating the inter-correlation between the 10 repetitions. Since number of synergies are two, 200 correlation processes are needed to identify the reference repetition which is achieves the highest average correlation coefficient between repetitions. 

The second step is to use this reference to arrange synergies within each repetition \cite{Barroso2014}. Finally, the arranged synergies are averaged to compute the first and second mean synergies for the task. Then, to identify the shared synergy of one \ac{dof}, the mentioned method is applied on the two tasks forming the \ac{dof} in question, the correlation coefficients between the resulting four mean synergies (two for each task) are calculated  so that  the highly correlated synergies between the two tasks are identified as shared, while the other two are considered as task-specific \cite{Torres-Oviedo2010,Chvatal2013a,Nazifi2017}.

\subsubsection{Comparison between shared synergies identified by \ac{ctd} and NMF}

We compared shared and task-specific synergies identified using the constrained Tucker tensor decomposition method with those identified by using the traditional \ac{NMF} and correlation method. This comparison is held  since there is no ground truth about the shared and task-specific synergies. Therefore, for each wrist's \ac{dof}, three synergies are identified by Tucker (two task-specific and one shared synergy) while four mean synergies (two for each task in the \ac{dof}) are estimated using \ac{NMF}. The correlation coefficient between Tucker and \ac{NMF} synergies are calculated and averaged across all 27 subjects. The comparison is held between the main three wrist's \acp{dof}: ulnar and radial deviation (DoF1); wrist extension/flexion  (DoF2); and wrist supination/pronation (DoF3).

\subsubsection{Validation with randomised repetitions}

In order to provide further validation to the approach of shared synergy identification using \ac{ctd}, the $repetition$ mode in the $3^{rd}$-order tensor of each \ac{dof} was randomly shuffled to destroy any task-repetition information. The same \ac{ctd} algorithm is applied on the tensor to identify the shared synergy between the two tasks. The 2 task-specific synergies will be corrupted since information about the tasks are missing. However, this experiment  tests the ability of constrained Tucker method to identify the shared synergies without any data arrangement which cannot be achieved using the traditional \ac{NMF} and correlation method. The shared synergies identified from the shuffled  $3^{rd}$-order tensors is compared against the shared synergies estimated from uncorrupted ones by calculating the correlation coefficients between them. The comparison is done using 15 shuffled tensors for each \ac{dof} of the main 3 wrist's \ac{dof} and the average correlated is computed.

\section{Results}

\subsection{Tensor decomposition for muscle synergy analysis}
	   \subsubsection{Tucker and PARAFAC models for synergy extraction} 

\begin{figure*}[t]
	\centering
	\begin{subfigure}[b]{0.9\textwidth}
		\centering
		\includegraphics[width=\linewidth]{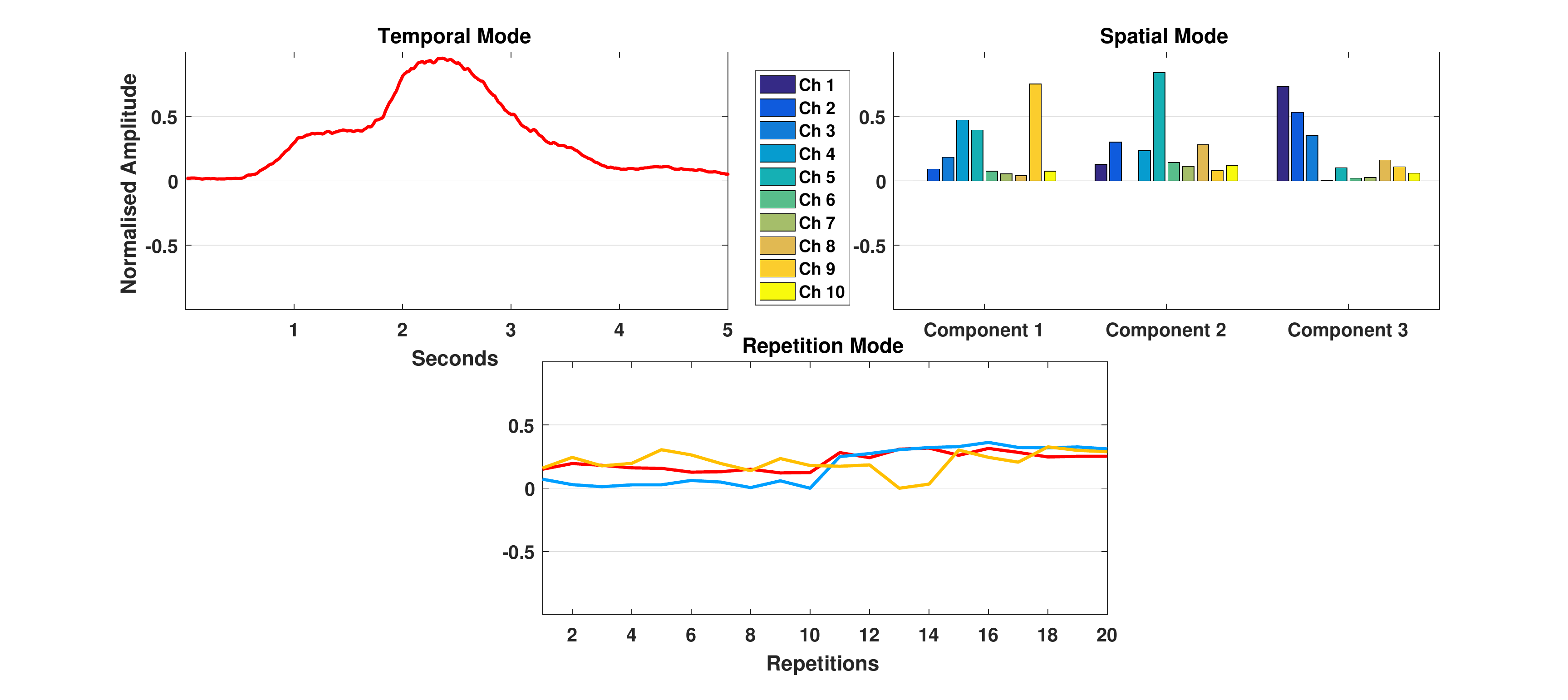}
		\caption{\textit{The average and standard deviation for 10 runs of non-negative $[3,3,3]$ Tucker decomposition for 1-DoF tensor.}}
		\label{fig:exampleTuckerNonNeg}
	\end{subfigure}
	\hfill
	\begin{subfigure}[b]{0.9\textwidth}
		\centering
		\includegraphics[width=\linewidth]{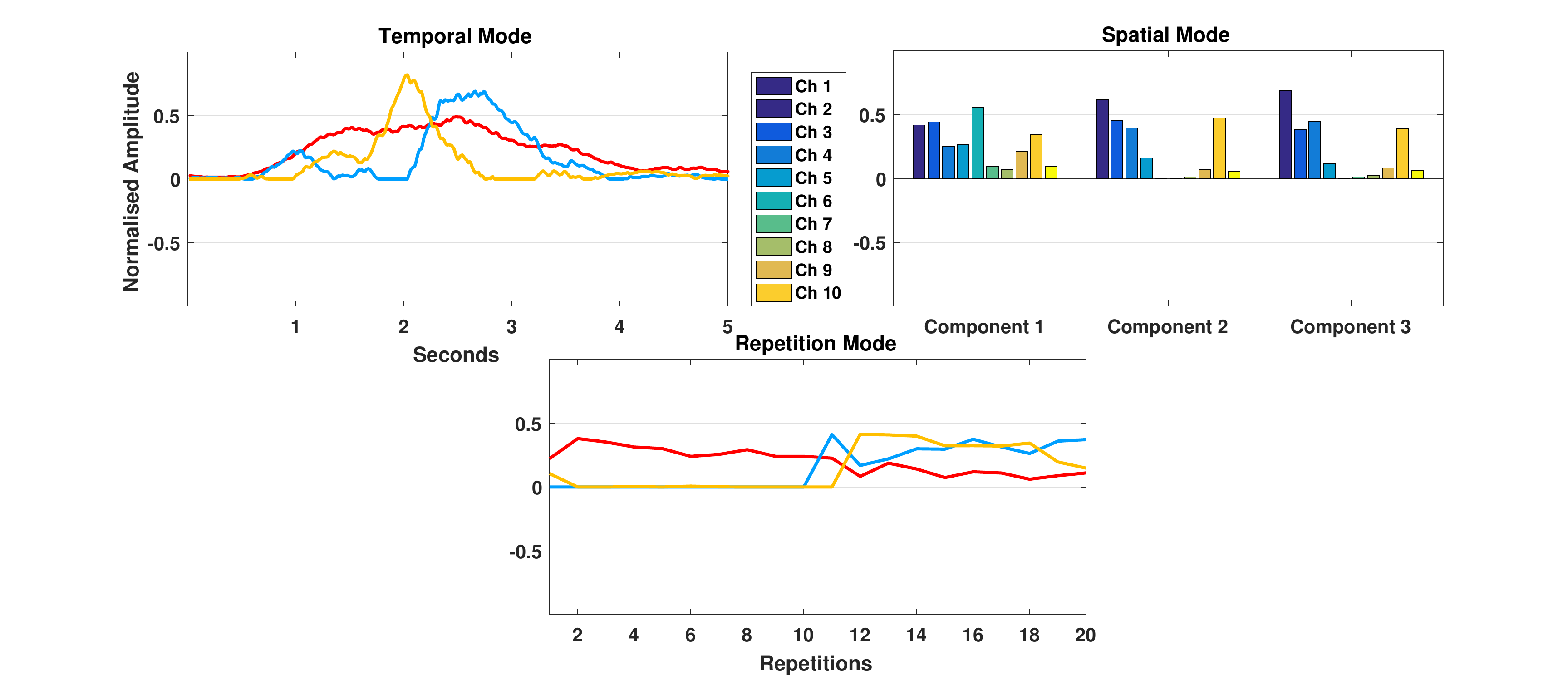}
		\caption{\textit{The average and standard deviation for 10 runs  of non-negative $3$-components PARAFAC decomposition for 1-DoF tensor.}}
		\label{fig:examplePARAFACNonNeg}
	\end{subfigure}
	\caption{The average (solid line) and standard deviations (shaded areas) for 10 runs of non-negative Tucker (\ref{fig:exampleTuckerNonNeg}) and  PARAFAC (\ref{fig:examplePARAFACNonNeg}) applied on the $3^{rd}$-order tensor. Because of the uniqueness of PARAFAC solution, its standard deviation is zero as shown in Panel \ref{fig:examplePARAFACNonNeg}. While only one component (blue) in Tucker seems to be unique in the $temporal$ and $repetition$ mode as shown in Panel \ref{fig:exampleTuckerNonNeg}.}
	\label{fig:examplestdParaandTucker}
\end{figure*}

Three non-negative Tucker decomposition models with $[2,2,2]$, $[3,3,3]$ and $[4,4,4]$ components were applied on 1- and 2-DoFs tensors for muscle synergy extraction.  An example of the 10 runs of the $[3,3,3]$ Tucker decomposition for 1-DoF tensor is shown in Figure~\ref{fig:exampleTuckerNonNeg}. The explained variance and the algorithm execution time were recorded for each decomposition and the median values across the 27 subjects are summarised in Table~\ref{table_Tucker}.

\begin{table}
\centering
\setlength{\extrarowheight}{0pt}
\addtolength{\extrarowheight}{\aboverulesep}
\addtolength{\extrarowheight}{\belowrulesep}
\setlength{\aboverulesep}{0pt}
\setlength{\belowrulesep}{0pt}
\caption{Median explained variance and execution time for the non-negative Tucker decomposition of the 27 Subjects.}
\label{table_Tucker}
\begin{tabular}{c|c|c|c|c} 
\toprule
 & \multicolumn{2}{c|}{\textbf{1-DoF Tensor} } & \multicolumn{2}{c}{\textbf{2-DoFs Tensor} } \\ 
\hline
\rowcolor[rgb]{0.937,0.937,0.937} \begin{tabular}[c]{@{}>{\cellcolor[rgb]{0.937,0.937,0.937}}c@{}} \textbf{No. of}\\\textbf{ Components} \end{tabular} & \begin{tabular}[c]{@{}>{\cellcolor[rgb]{0.937,0.937,0.937}}c@{}}Explained\\ Variance\end{tabular} & time(s) & \begin{tabular}[c]{@{}>{\cellcolor[rgb]{0.937,0.937,0.937}}c@{}}Explained\\ Variance\end{tabular} & time(s) \\ 
\hline
\begin{tabular}[c]{@{}c@{}}\textbf{[2,2,2]}\\ \end{tabular} & 87.8\% & 12.5 & 77.5\% & 24.9 \\ 
\hline
\textbf{[3,3,3]} & 92.2\% & 25.7 & 86.4\% & 59.7 \\ 
\hline
\begin{tabular}[c]{@{}c@{}}\textbf{[4,4,4]}\\ \end{tabular} & 94.3\% & 73 & 89.8\% & 75.2 \\
\bottomrule
\end{tabular}
\end{table}

The \ac{para} decomposition model was applied on both 1- and 2-DoFs tensors of wrist horizontal and vertical \acp{dof}. The number of components explored were 2, 3 and 4 where a non-negativity constraint was applied on all components. An example of 3-component \ac{para} decomposition on 2-DoFs tensor is shown in Figure~\ref{fig:examplePARAFACNonNeg}. The time of execution for \ac{para} algorithm as well as \ac{corc} were recorded across the 27 subjects are summarised in Table~\ref{table_para}.

\begin{table}
\centering
\setlength{\extrarowheight}{0pt}
\addtolength{\extrarowheight}{\aboverulesep}
\addtolength{\extrarowheight}{\belowrulesep}
\setlength{\aboverulesep}{0pt}
\setlength{\belowrulesep}{0pt}
\caption{The median core consistency and time of execution for the PARAFAC decomposition across the 27 subjects.}
\label{table_para}
\begin{tabular}{c|c|c|c|c} 
\toprule
 & \multicolumn{2}{c|}{\textbf{1-DoF Tensor} } & \multicolumn{2}{c}{\textbf{2-DoFs Tensor} } \\ 
\hline
\rowcolor[rgb]{0.937,0.937,0.937} \begin{tabular}[c]{@{}>{\cellcolor[rgb]{0.937,0.937,0.937}}c@{}} \textbf{No. of}\\\textbf{ Components} \end{tabular} & \begin{tabular}[c]{@{}>{\cellcolor[rgb]{0.937,0.937,0.937}}c@{}}Core\\consistency \end{tabular} & time(s) & \begin{tabular}[c]{@{}>{\cellcolor[rgb]{0.937,0.937,0.937}}c@{}}Core\\consistency \end{tabular} & time(s) \\ 
\hline
\begin{tabular}[c]{@{}c@{}}\textbf{2}\\ \end{tabular} & 95.2\%  & 0.39 & 91.1\%  & 0.58 \\ 
\hline
\textbf{3}  & 30\%~ & 0.60 & 64.6\% & 0.72 \\ 
\hline
\begin{tabular}[c]{@{}c@{}}\textbf{4}\\ \end{tabular} & 6\%  & 0.91 & 29.3\% & 1.13 \\
\bottomrule
\end{tabular}
\end{table}

\subsubsection{Constrained Tucker decomposition}

\begin{figure*}[t]
	\centering
	\includegraphics[width=\textwidth]{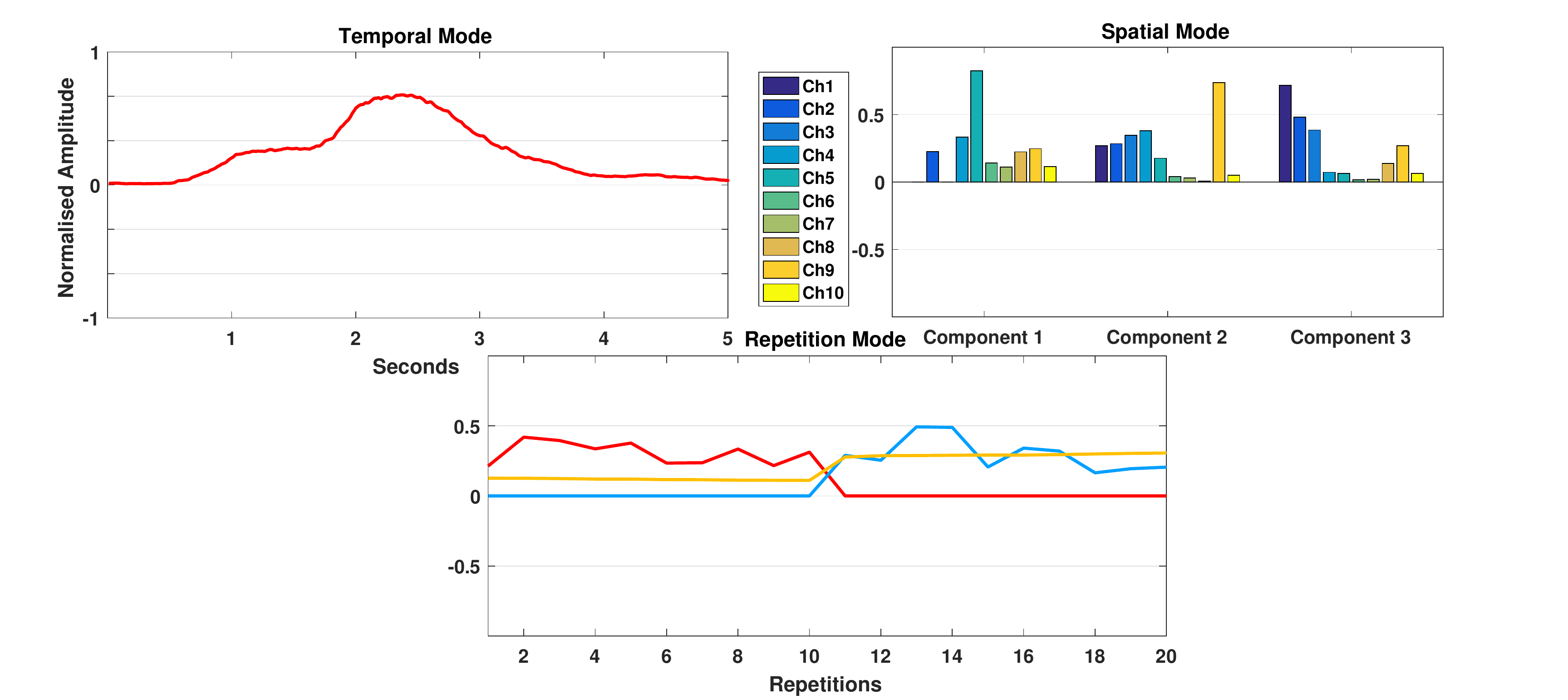}
	\caption{Constrained $[1,3,3]$ Tucker decomposition for the $3^{rd}$-order tensor in Fig. (\ref{fig:exampleTensor}). The $spatial$ mode had 3 components, the first 2 components are muscle synergies specified  for the ulnar and radial deviation movements while the third synergy represents the shared synergy between them.}
	\label{fig:TuckerConstraint3}
\end{figure*}

The \ac{ctd} models were applied on 1- and 2-DoFs tensors for muscle synergy estimation for 10 runs across the 27 subjects. The 1-DoF tensor was decomposed using $[1,3,3]$ constrained Tucker method  while the 2-DoFs tensor was decomposed using $[2,5,5]$ constrained Tucker model.  An example of $[1,3,3]$ constrained Tucker method for Tensor shown in Figure~\ref{fig:exampleTensor} is illustrated in Figure~\ref{fig:TuckerConstraint3}. Explained variance and execution time were recorded and the median values are shown in Table~\ref{table_ConstTucker}.

\begin{table}
\centering
\setlength{\extrarowheight}{0pt}
\addtolength{\extrarowheight}{\aboverulesep}
\addtolength{\extrarowheight}{\belowrulesep}
\setlength{\aboverulesep}{0pt}
\setlength{\belowrulesep}{0pt}
\caption{The median explained variance and time for execution for the \ac{ctd} across the 27 Subjects.}
\label{table_ConstTucker}
\begin{tabular}{c|c|c} 
\toprule
 & \textbf{1-DoF Tensor}  & \textbf{2-DoF Tensor}  \\ 
\hline
\rowcolor[rgb]{0.937,0.937,0.937} \begin{tabular}[c]{@{}>{\cellcolor[rgb]{0.937,0.937,0.937}}c@{}}\textbf{ No. of Components}\\ \end{tabular} & [1,3,3] & [2,5,5] \\ 
\hline
\textbf{Explained Variance}  & 78.28\% & 73.21\% \\ 
\hline
\textbf{~time(s)} & 0.26 & 0.65 \\
\bottomrule
\end{tabular}
\end{table}

\subsection{Shared synergy identification}

\subsubsection{NMF synergies}

\begin{figure}[ht]
	\centering

	\begin{subfigure}{0.49\textwidth}
		\centering
		\includegraphics[width=\linewidth]{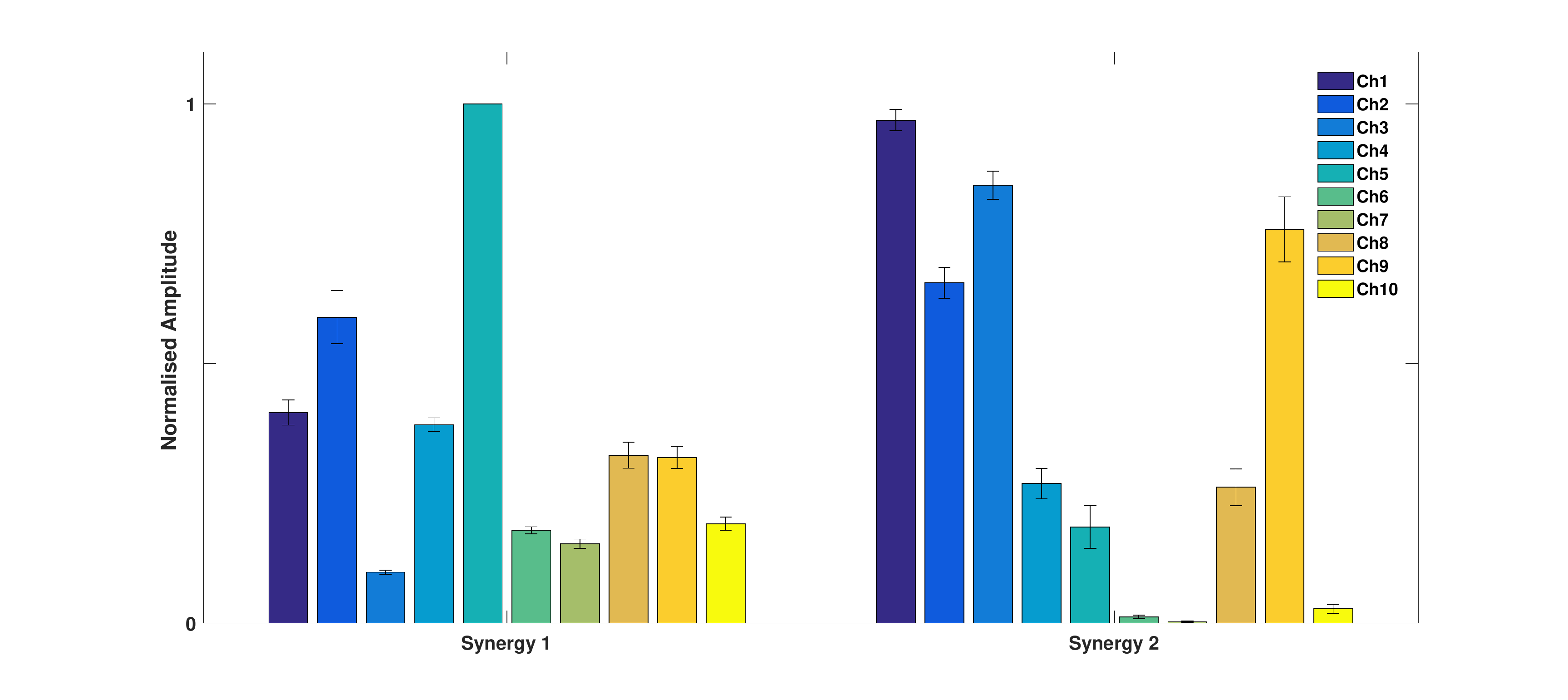}
		\caption{}
		\label{fig:nmftask16}
	\end{subfigure}
	\hfill
	\begin{subfigure}{0.49\textwidth}
		\centering
		\includegraphics[width=\linewidth]{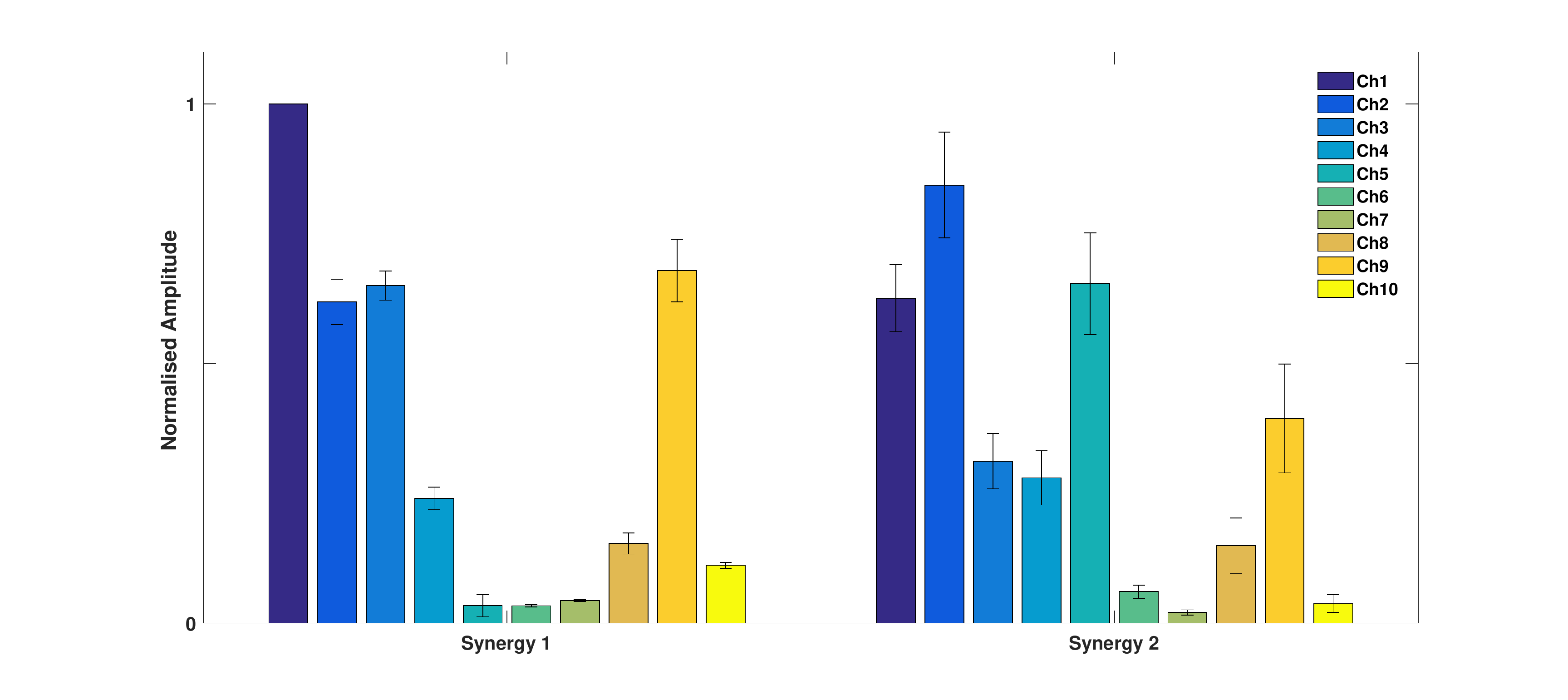}
		\caption{}
		\label{fig:nmftask17}
	\end{subfigure}

\caption{ two-component NMF for the ulnar (\ref{fig:nmftask16}) and radial (\ref{fig:nmftask17}) deviations movements of subject ``1" averaged across 10 repetitions for each task. The second component of ulnar deviation is highly correlated with the first component of radial deviation suggesting that these are the shared synergies between those tasks. }
	\label{fig:Nmftask16and17}
\end{figure}
 
A number of wrist tasks were selected and 10-channel \ac{EMG} recording was decomposed using \ac{NMF} to extract two synergies for each task. Our analysis found that two synergies could account for over  90\% of the variability in data for all repetitions. For each task, \ac{NMF} was applied on each of the 10 repetitions and the estimated synergies were rearranged using mutual correlation coefficients then averaged across repetitions to result two muscle synergies for each movement. An example of the averaged synergies are shown in Figure \ref{fig:Nmftask16and17} for the ulnar and radial deviation movements (DoF1) of subject ``1".
 
Shared synergies are determined through correlation. As shown in Figure \ref{fig:Nmftask16and17}, the second synergy of ulnar deviation (Fig.~\ref{fig:nmftask16}) is highly correlated with the first component of radial deviation  (Fig.~\ref{fig:nmftask17}) with $r=0.91$. Therefore, according to the standard \ac{NMF} approach the average of these two synergies is considered as a shared synergy between the ulnar and radial deviation tasks while the remaining synergies are task-specific.

\subsubsection{Shared synergies comparison}

\begin{figure}[ht]
	\centering
	
	\begin{subfigure}{0.5\textwidth}
		\centering
		\includegraphics[width=\textwidth]{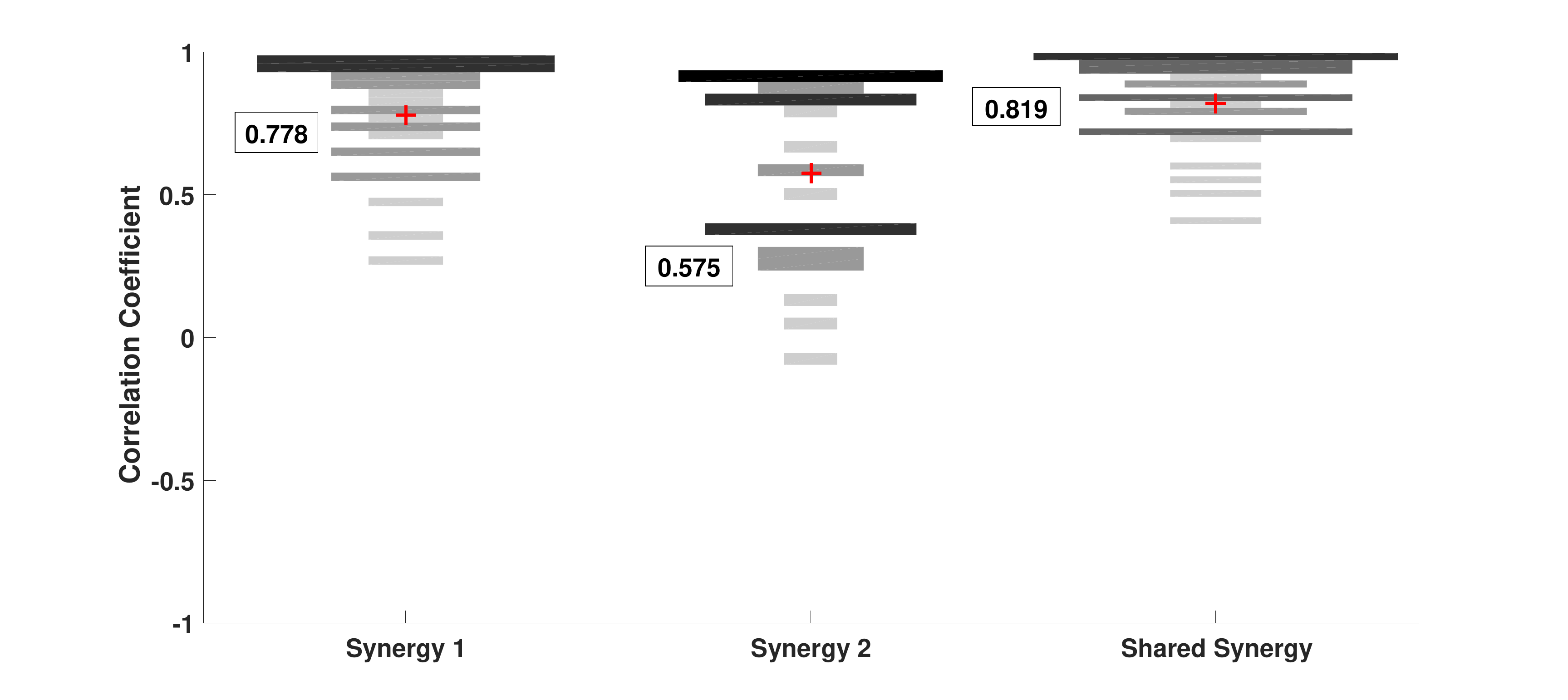}
		\caption{}
		\label{fig:CorrViolinTask16NMF2}
	\end{subfigure}
	
	\hfill
	
	\begin{subfigure}{0.5\textwidth}
		\centering
		\includegraphics[width=\textwidth]{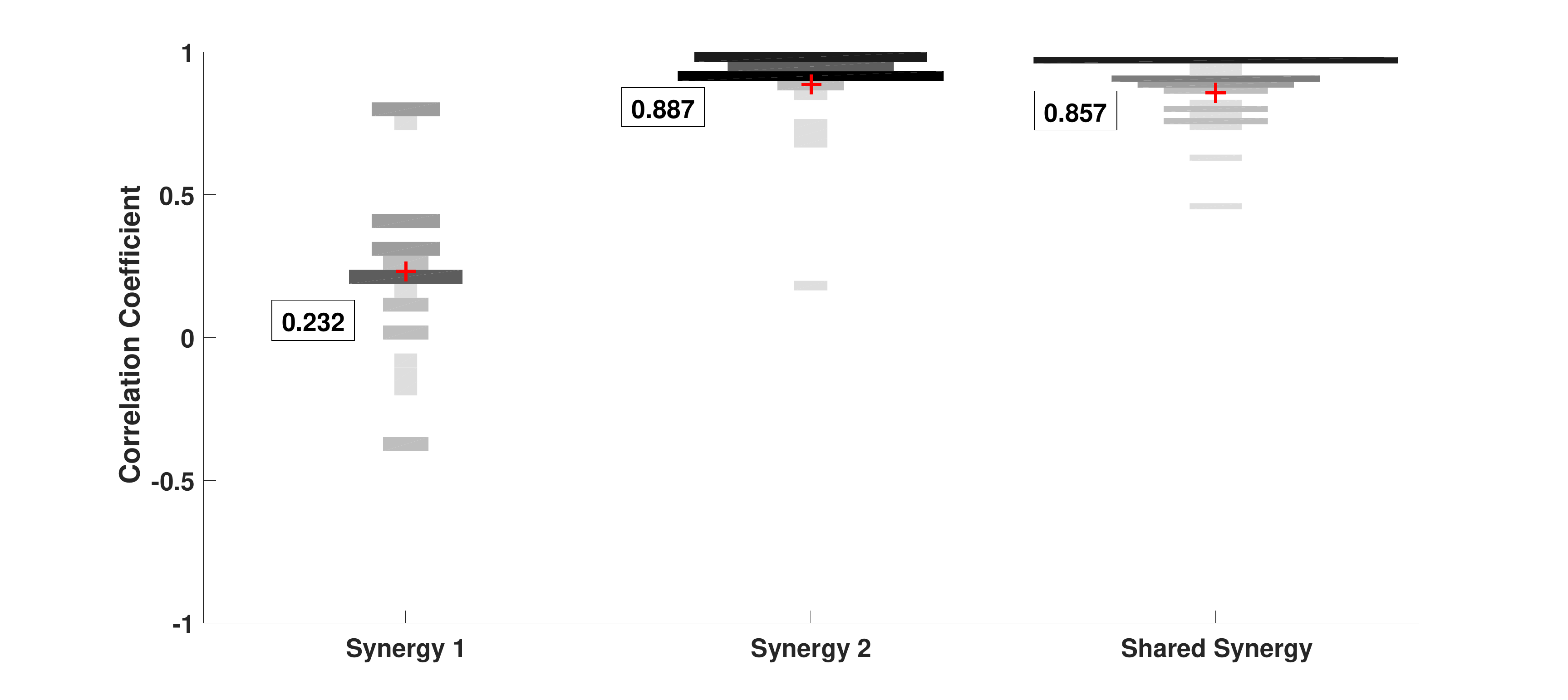}
		\caption{}
		\label{fig:CorrViolinTask17NMF2}
	\end{subfigure}
	
	\caption{Visualisation of six histograms for the correlation coefficients between synergies extracted by \ac{ctd} and  NMF  synergies across the 27 subjects for ulnar (Panel \ref{fig:CorrViolinTask16NMF2})  and  radial (Panel~\ref{fig:CorrViolinTask17NMF2}) deviation (DoF1). Each line represents  frequency of occurrence for the histogram, where darker shades refer to higher frequency of occurrence. The red crosses each histogram are its mean value. The full mean value comparison across all DoFs is represented in  Table~\ref{table_results}.}

	\label{fig:Results&Stats}
\end{figure}

Synergies extracted from \ac{ctd} are compared against \ac{NMF} synergies to test the ability of this method to identify shared and task-specific synergies. The correlation coefficients between Tucker and \ac{NMF} synergies are used as a metric. This was done for 3 pairs of tasks (\acp{dof}) for the wrist movements across the 27 subjects in the dataset, where the 3 synergies from Tucker decomposition are compared against each averaged \ac{NMF} synergies of each task. For example, the correlation coefficients  between the estimated \ac{ctd} synergies and \ac{NMF} synergies of ulnar and radial deviation (\ac{dof}) for the 27 subjects are represented in Figure~\ref{fig:Results&Stats}. The average correlations for all 3 DoFs are summarised in Table~\ref{table_results}.

The third (shared) synergy is highly correlated with both tasks as the average correlation coefficients for ulnar and radial deviation are $0.819$ and $0.857$, respectively. Each of the other 2 task-specific synergies are correlated with its respective task. For ulnar deviation, the first synergy has correlation coefficient of $0.778$ compared to $0.575$ for the second synergy, while for the radial deviation the second synergy has an average correlation coefficient of $0.887$ compared to $0.232$ with the first synergy. Similar results  are found with other movements such as wrist extension/flexion and  wrist supination/pronation) as shown in Table \ref{table_results}.

\begin{table}
\centering
\setlength{\extrarowheight}{0pt}
\addtolength{\extrarowheight}{\aboverulesep}
\addtolength{\extrarowheight}{\belowrulesep}
\setlength{\aboverulesep}{0pt}
\setlength{\belowrulesep}{0pt}
\caption{Average correlation coefficients between Tucker and NMF synergies for the 3 Main DoFs of wrist. }
\label{table_results}
\begin{tabular}{c|c|c|c|c} 
\toprule
\rowcolor[rgb]{0.937,0.937,0.937}  &  & Synergy 1 & Synergy 2 & Synergy 3 \\ 
\hline
\multirow{2}{*}{DOF 1} & Ulnar deviation & 0.778 & 0.575 & 0.819 \\
 & Radial deviation & 0.232 & 0.887 & 0.857 \\ 
\hline
\rowcolor[rgb]{0.937,0.937,0.937} {\cellcolor[rgb]{0.937,0.937,0.937}} & Wrist extension & 0.729 & 0.337 & 0.868 \\
\rowcolor[rgb]{0.937,0.937,0.937} \multirow{-2}{*}{{\cellcolor[rgb]{0.937,0.937,0.937}}DOF 2} & Wrist flexion & 0.408 & 0.776 & 0.880 \\ 
\hline
\multirow{2}{*}{DOF 3} & Wrist supination & 0.911 & 0.481 & 0.879 \\
 & Wrist pronation & 0.104 & 0.920 & 0.792 \\
\bottomrule
\end{tabular}
\end{table}

\subsubsection{Validation with randomised repetitions}

\begin{figure*}[t]
	\centering
	\includegraphics[width=\textwidth]{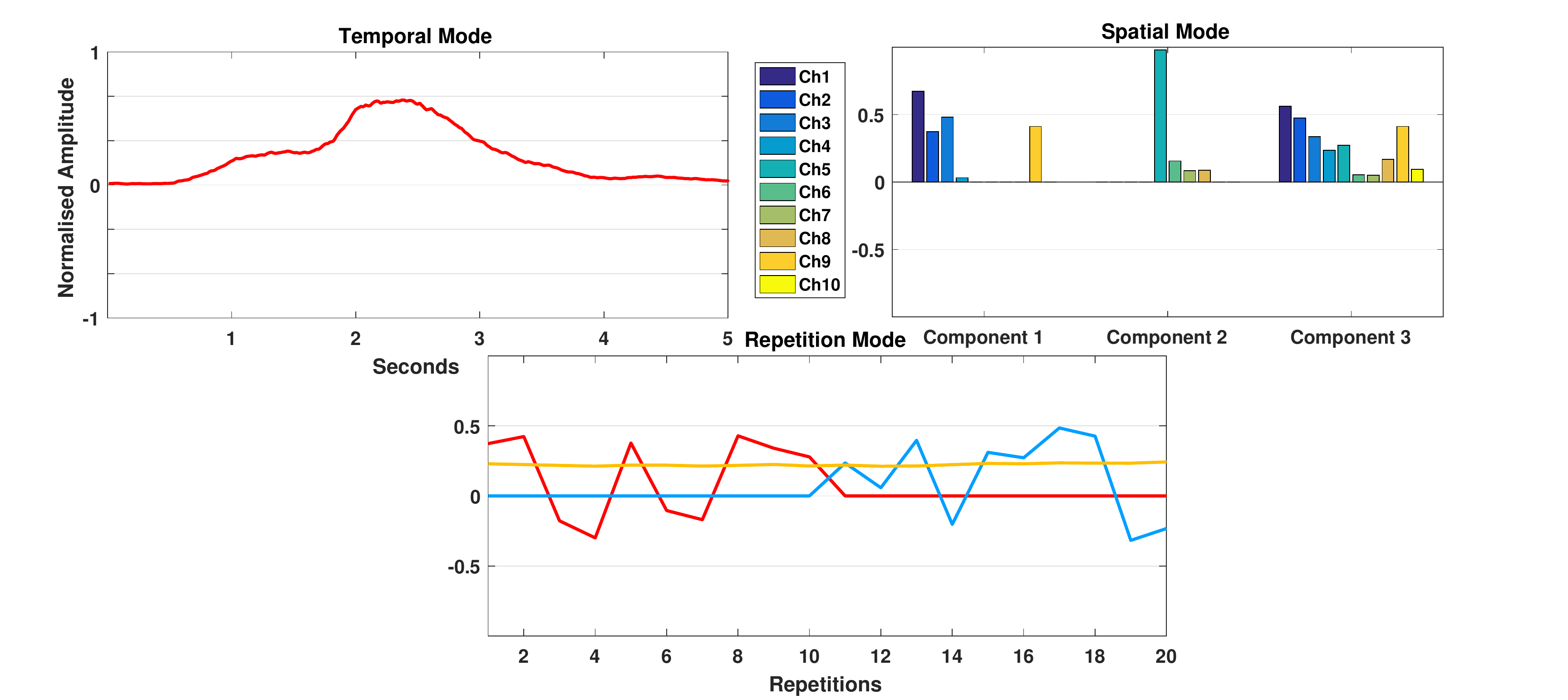}
	\caption{\ac{ctd} for the DoF1 tensor in Figure (\ref{fig:exampleTensor}) but with shuffled $repetition$ mode. The algorithm was able to identify the same shared synergy (third $spatial$ component) as the decomposition of the regular tensor (Fig. \ref{fig:TuckerConstraint3}) even without any task-repetition information. }
	\label{fig:randomtucker3}
\end{figure*}

In order to validate the approach of shared synergy identification and to show that it is robust to any repetitions disarrangement, the $3^{rd}$-order tensor in  Figure \ref{fig:exampleTensor}  was randomly shuffled across the $repetition$ mode to destroy the task-repetition information. The \ac{ctd}  was applied on the randomly shuffled tensor to identify shared synergy as shown in Figure \ref{fig:randomtucker3}. In comparison with the normal tensor decomposition (Fig.~\ref{fig:TuckerConstraint3}), we noticed that the task-specific components were different as expected since the information was destroyed. On the other hand,  the shared synergy in the $spatial$ mode were very similar. The average correlation coefficients between shared synergies identified from 15 shuffled tensors and  from arranged ones were found to be $0.89$. This shows the ability of the algorithm to identify the shared synergy despite the corruption in the task-repetition information during the tensor construction.

\section{Discussion}

We proposed the use of higher-order tensors and their decompositions as a framework for muscle synergy analysis. Although it may not fully agree with the classical definition of muscle synergies, it is inspired by it to provide more synergistic information from the data. This is motivated by the fact that most of \ac{EMG} datasets are naturally in multi-way form with different repetitions from tasks and/or subjects. In addition, some synergy analysis techniques \cite{Delis2014,Semprini2017,Delis2015} extracted spatial and temporal components of muscle activity using a space-by-time decomposition model that resembles Tucker2 tensor decomposition model. Surprisingly, a fully developed tensor factorisation has barely been widely used in \ac{EMG} analysis \cite{Ebied2017}.
 
The main objective was to explore the use of tensor factorisation models in muscle synergy analysis. We proposed a \ac{ctd} model inspired by the shared synergy concept and compare it with the most prominent methods (\ac{para} and Tucker decomposition models). This was approached by applying these methods with different number of components on $3^{rd}-$order tensors consisting of 20 and 40 multichannel \ac{EMG} repetitions for one and two wrist's \acp{dof} respectively (10 repetitions each). The three tensor decomposition methods were assessed according to the algorithm execution time, explained variance and \ac{corc}, as appropriate. However, we acknowledge that execution time is for comparison reasons only, not for assessing the whole performance. The constraint Tucker decomposition model was the best approach to extract muscle synergies from $3^{rd}$-order tensor as it was capable to estimate unique synergies in a short execution time with acceptable explained variance. 

The secondary objective was to illustrate the potential use of tensor decomposition  for shared and task-specific synergies identification. The proposed \ac{ctd} method was used to identify shared synergies between each pair of tasks that forms a main \ac{dof} of wrist movements. The resulting synergies were compared against the standard \ac{NMF} factorisation approach for the 3 wrist \acp{dof} across 27 subjects. The results showed that shared and task-specific synergies estimated via the \ac{ctd} method were highly correlated with those identified through traditional \ac{NMF} approach.  In addition, tensor shared synergies were compared to randomly shuffled tensor without any task-repetition information for further validation and the proposed algorithm was able to estimate nearly the same shared synergy as the ordered tensor, thus showing robustness to disarrangement. 

\subsection{Tensor decomposition models for muscle synergy analysis}

The comparison between Tucker, \ac{para} and \ac{ctd}  models for muscle synergy analysis showed that Tucker decomposition can provide a good fit for the data as shown by the high explained variance percentage in Table~\ref{table_Tucker}. However, the estimated synergies via non-negative Tucker decomposition were inconsistent as shown in Figure~\ref{fig:exampleTuckerNonNeg}. This inconsistency arises from the fact that Tucker decomposition generally does not provide unique solutions \cite{Cichocki2014}, and uniqueness can be achieved in practice by imposing additional constraints \cite{Zhou2012}. Since the decomposition is unconstrained, the initialisation (which is set randomly) changes in each run. This would result in different components in each run since there are no constraints to achieve uniqueness of the solution and the model converges to different local minima. This also explain the longer execution time since the unconstrained Tucker algorithm took more iterations to converge takes more time to converge as represented in Table~\ref{table_Tucker}. Despite the increase of explained variance percentage with the additional number of components, the execution time increased as well, since the algorithm could not converge easily. 

On the other hand, \ac{para} was significantly faster as seen in Table~\ref{table_para}, since it converged to the same local minima most of the time due to its extreme constrained nature. \ac{para} with non-negativity constraints was capable of estimating estimate muscle synergies from the 1-DoF tensor as shown in Figure~\ref{fig:examplePARAFACNonNeg}. However,  \ac{para} could not deal with 2-DoFs tensors or higher number of synergies as the decomposition deviates from the trilinear model and \ac{para} could not hold. This is illustrated with low \ac{corc} as shown in Table~\ref{table_para}. In addition, synergy estimation is affected by inflexibility of the \ac{para} model as the number of components are fixed across modes. Therefore, the  information of each task in the $temporal$ mode is segmented between components.

Two \ac{ctd} models ($[1,3,3]$ and $[2,5,5]$) were proposed to decompose the 1-\ac{dof} and 2-\acp{dof} tensors respectively. They were able to achieve over 70\% explained variance and decrease the execution time by about 10-fold compared to the non-negative Tucker model as shown in Table~\ref{table_ConstTucker}. Moreover, the resulting synergies were consistent over the runs as shown in Figure~\ref{fig:TuckerConstraint3} unlike Tucker model. The \ac{ctd} approach allocates one synergy for each movement and additional ```shared" synergy to account for variability inspired by the shared synergy concept. This additional shared synergy improved the  explained variance compared to $[1,2,2]$ constrained Tucker model where the median explained variance was $59.3\%$ in the preliminary results.

Moreover, the total number of components in \ac{ctd} may be greater than or equal to the number of components in traditional Tucker decomposition. However, the total number of elements for \ac{ctd} that need to be estimated is significantly less than Tucker model. For example, the 1-\ac{dof} tensor ($500 sample \times 10 channels \times 20 repetitions$) with number of elements=$100,000$ is decomposed via $[2,2,2]$ Tucker decomposition into $1060$ elements in addition to $8$ elements in the core tensor. On the other hand, a $[1,3,3]$ \ac{ctd} can decompose the same tensor into $590$ elements in addition to $9$ elements in its sparse core tensor. Hence, the proposed \ac{ctd} for synergy extraction is more efficient in comparison to unconstrained Tucker model.  

Two important variants for the Tucker decomposition models are worth noting, Tucker1 and Tucker2, which can be seen as a special case of Tucker model where only one and two modes are estimated respectively. In both models, the additional modes are set to be identity matrices and absorbed into the core tensor \cite{Mrup2011b}. As a result, Tucker1 model is equivalent to the ordinary two-dimensional PCA, while Tucker2 is a model of intermediate complexity as compared with the Tucker1 and the standard Tucker model \cite{Kolda2008c}. In Tucker2, the third mode is absorbed to the core tensor and the model explains the variability of the first two modes only. Hence, Delis \textit{el al.}~\cite{Delis2014} approach focused on \textit{spatial} and \textit{temporal} components and their interactions, while the \ac{ctd} method adds \textit{repetition} mode to \textit{spatial} and \textit{temporal} modes to incorporate different movements and the shared synergies  between them to provide a 3$^{rd}$-order   \ac{EMG} tensor decomposition for muscle synergy analysis.

Hence, we conclude that the proposed \ac{ctd} method is the best solution to obtain unique and interpretable synergies from 3$^{rd}$-order tensor decomposition. It was capable to achieve this with smallest number of synergies and elements because of the utilisation of shared synergy concept. The non-negativity constraint is essential because of the additive nature of synergies. Moreover, the fixed core tensor is pivotal, as we can directly relate synergies (in the $spatial$ mode) to other specific components in $temporal$ and $repetition$ modes. This is in contrary with  the unconstrained  core tensor in Tucker model, which allows for interactions between all components in each mode. Due to these interactions, it becomes difficult to achieve a unique of solution for Tucker decomposition. This increases the computational time dramatically.

\subsection{Shared muscle synergy identification}

 We showed that higher-order tensor decomposition models can achieve direct identification of shared synergies without relying on any similarity metric such as correlation coefficient. This is in contrast with the current approaches for shared synergy estimation \cite{Nazifi2017,Martino2015,Barroso2014,Chvatal2013a,Chvatal2011a} which apply \ac{NMF} repetitively on multi-channel EMG recordings of different repetitions and then rely on maximising the correlation coefficients between the estimated synergies with regard to a reference one. Then, shared and task-specific synergies are identified through the correlation coefficient threshold.  This was illustrated in Figure~\ref{fig:TuckerConstraint3} of \ac{ctd} model, where component 1 and 2 in the $spatial$ mode are task-specific for the two task forming the tensor while component 3 is the shared synergy between them.

Synergies identified via \ac{ctd} were compared  against  synergies extracted using \ac{NMF} for the 27 subjects. In spite of the potential drawbacks of \ac{NMF} shared synergies, we used \ac{NMF} as benchmark since there is no ground truth for shared synergies to compare both methods against. In addition, the wrist movement included in the study are limited  since shared synergies are easier to identify. Only two \ac{NMF} synergies could explain over 90\% of variance. Hence, the errors of disarrangement is minimised. This was done for 3 pairs of tasks (\acp{dof}) for the wrist movements (Table~\ref{table_results}). 

The shared synergies identified by Tucker (third synergy in Figure~\ref{fig:TuckerConstraint3}) were highly correlated with both tasks while each of the other two tasks correlated with one task as a task-specific synergy. This highlights the ability of \ac{ctd} to identify task-specific and shared synergies directly from the multi-way datasets. Further validation were held by applying the \ac{ctd} on randomly shuffled tensor without any task-repetition information. The proposed algorithm was able to estimate nearly the same shared synergy as the ordered tensor (Fig.~\ref{fig:randomtucker3}), which indicated robustness of the method. In addition, the standard \ac{NMF} approaches for shared and task-specific synergies identification are vulnerable to errors and biases since they depend on the particular arrangement of the data, the choice of the reference synergy, and the correlation coefficient threshold. This is not the case for \ac{ctd} approach where it was able to identify the shared synergy even with a shuffled tensor as shown in Figure~\ref{fig:randomtucker3}. In addition, it is a more direct and faster alternative since there is no need to apply repetitive \ac{NMF} and correlation.

 \subsection{Applications, limitations and future work}

Usually, the shared and task-specific synergies are identified in a study of complex multi-joint movements such as gait and posture analysis \cite{Nazifi2017,Martino2015,Barroso2014}. However, we choose simple wrist movements for few reasons. Firstly, this is a first study to show how higher-order tensors could be beneficial for muscle synergy analysis. Secondly, wrist movements are simple tasks that could be described by two synergies as mentioned before. Therefore, shared synergies can be identified easily with minimum disarrangement errors for good comparison and validation for our proposed tensor approach. Finally, we are interested in upper-limb myoelectric control and looking to the shared synergy concept as an inspiration for proportional myoelectric control based on muscle synergies in the future.

Moreover, the main aim for this is study is to highlight the potential of higher-order tensor model for muscle activity analysis especially extracting muscle synergies. Hence, \ac{ctd} could be extended to various applications by converting the information we have into the right set of constraints. For example, this approach could be extended to estimate the shared synergies across subjects to explore the subject-specific synergies \cite{Torres-Oviedo2010}. In addition, and in relation to the point above, different set of constraints could help to develop a myoelectric control based on muscle synergies as in \cite{Jiang2014a}.

\section{Conclusion}

In conclusion, we introduced tensor decomposition models (\ac{para} and Tucker) for muscle synergy extraction and compared their use in \ac{EMG} analysis to extract meaningful muscle synergies with a proposed \ac{ctd} model. The developed method was the best approach for muscle synergy estimation by providing unique and interpretable synergies with high explained variance and short execution time. The proposed \ac{ctd} model can be used to identify shared and task-specific synergies. The results were compared against the standard \ac{NMF} approach using data from the publicly available Ninapro dataset. The \ac{ctd} method was more suitable to the multi-way nature of the datasets without relying on symmetry metrics or synergies arrangements. Furthermore, it provided more direct and data-driven estimations of the synergies in comparison with \ac{NMF}-based approaches, making our approach more robust to disarrangement of repetitions and the loss of task-repetition information. Thus, we expect that this study will pave the way for the development of muscle activity processing and analysis methods based on higher-order techniques.


\bibliographystyle{ieeetr}
\bibliography{mendeley_v2}

\begin{thebibliography}{10}

\bibitem{Cichocki2014}
A.~Cichocki, D.~Mandic, A.~H. Phan, C.~Caiafa, G.~Zhou, Q.~Zhao, and
  L.~De~Lathauwer, ``{Tensor Decompositions for Signal Processing Applications
  From Two-way to Multiway Component Analysis},'' {\em IEEE Signal Processing
  Magazine}, vol.~32, pp.~1--23, Mar. 2014.

\bibitem{Tucker1966a}
L.~R. Tucker, ``{Some mathematical notes on three-mode factor analysis},'' {\em
  Psychometrika}, vol.~31, pp.~279--311, Sept. 1966.

\bibitem{Harshman1970a}
R.~a. Harshman, ``{Foundations of the PARAFAC procedure: Models and conditions
  for an {\textquotedblleft}explanatory{\textquotedblright} multimodal factor
  analysis},'' {\em UCLA Working Papers in Phonetics}, vol.~16, no.~10,
  pp.~1--84, 1970.

\bibitem{Cong2015}
F.~Cong, Q.-H. Lin, L.-D. Kuang, X.-F. Gong, P.~Astikainen, and T.~Ristaniemi,
  ``{Tensor decomposition of EEG signals: A brief review},'' {\em Journal of
  Neuroscience Methods}, vol.~248, pp.~59--69, June 2015.

\bibitem{Spyrou2016}
L.~Spyrou, S.~Kouchaki, and S.~Sanei, ``{Multiview Classification and
  Dimensionality Reduction of Scalp and Intracranial EEG Data through Tensor
  Factorisation},'' Aug. 2016.

\bibitem{Escudero2015}
J.~Escudero, E.~Acar, A.~Fern{\'{a}}ndez, and R.~Bro, ``{Multiscale entropy
  analysis of resting-state magnetoencephalogram with tensor factorisations in
  Alzheimer's disease},'' {\em Brain Research Bulletin}, vol.~119,
  pp.~136--144, 2015.

\bibitem{Ebied2017}
A.~Ebied, L.~Spyrou, E.~Kinney-Lang, and J.~Escudero, ``{On the use of
  higher-order tensors to model muscle synergies},'' in {\em 2017
  39\textsuperscript{th} Annual International Conference of the IEEE
  Engineering in Medicine and Biology Society (EMBC)}, pp.~1792--1795, IEEE,
  July 2017.

\bibitem{Xie2013a}
P.~Xie and Y.~Song, ``{Multi-domain feature extraction from surface EMG signals
  using nonnegative tensor factorization},'' in {\em 2013 IEEE International
  Conference on Bioinformatics and Biomedicine}, pp.~322--325, IEEE, Dec. 2013.

\bibitem{Delis2014}
I.~Delis, S.~Panzeri, T.~Pozzo, and B.~Berret, ``{A unifying model of
  concurrent spatial and temporal modularity in muscle activity},'' {\em
  Journal of Neurophysiology}, vol.~111, no.~3, pp.~675--693, 2014.

\bibitem{Delis2015}
I.~Delis, S.~Panzeri, T.~Pozzo, and B.~Berret, ``{Task-discriminative
  space-by-time factorization of muscle activity},'' {\em Frontiers in Human
  Neuroscience}, vol.~9, p.~399, July 2015.

\bibitem{Hilt2018}
P.~M. Hilt, I.~Delis, T.~Pozzo, and B.~Berret, ``{Space-by-Time Modular
  Decomposition Effectively Describes Whole-Body Muscle Activity During Upright
  Reaching in Various Directions},'' {\em Frontiers in Computational
  Neuroscience}, vol.~12, p.~20, Apr. 2018.

\bibitem{Tresch1999}
M.~C. Tresch, P.~Saltiel, and E.~Bizzi, ``{The construction of movement by the
  spinal cord.},'' {\em Nature neuroscience}, vol.~2, pp.~162--7, Feb. 1999.

\bibitem{Saltiel2001}
P.~Saltiel, K.~Wyler-Duda, A.~D'Avella, M.~C. Tresch, and E.~Bizzi, ``{Muscle
  synergies encoded within the spinal cord: evidence from focal intraspinal
  NMDA iontophoresis in the frog.},'' {\em Journal of neurophysiology},
  vol.~85, pp.~605--619, Feb. 2001.

\bibitem{DAvella2003}
A.~d'Avella, P.~Saltiel, and E.~Bizzi, ``{Combinations of muscle synergies in
  the construction of a natural motor behavior.},'' {\em Nature neuroscience},
  vol.~6, pp.~300--308, Mar. 2003.

\bibitem{DAvella2015}
A.~d'Avella, M.~Giese, Y.~P. Ivanenko, T.~Schack, and T.~Flash, ``{Editorial:
  Modularity in motor control: from muscle synergies to cognitive action
  representation.},'' {\em Frontiers in computational neuroscience}, vol.~9,
  p.~126, Jan. 2015.

\bibitem{Cheung2012b}
V.~C.-K.~K. Cheung, A.~Turolla, M.~Agostini, S.~Silvoni, C.~Bennis, P.~Kasi,
  S.~Paganoni, P.~Bonato, and E.~Bizzi, ``{Muscle synergy patterns as
  physiological markers of motor cortical damage},'' {\em Proceedings of the
  National Academy of Sciences}, vol.~109, pp.~14652--14656, Sept. 2012.

\bibitem{Kutch2012}
J.~J. Kutch and F.~J. Valero-Cuevas, ``{Challenges and new approaches to
  proving the existence of muscle synergies of neural origin.},'' {\em PLoS
  computational biology}, vol.~8, p.~e1002434, Jan. 2012.

\bibitem{Bizzi2013}
E.~Bizzi and V.~C.-K.~K. Cheung, ``{The neural origin of muscle synergies.},''
  {\em Frontiers in computational neuroscience}, vol.~7, p.~51, Feb. 2013.

\bibitem{TRESCH}
M.~C. Tresch and A.~Jarc, ``{The case for and against muscle synergies},'' {\em
  Current Opinion in Neurobiology}, vol.~19, pp.~601--607, Dec. 2009.

\bibitem{Pons2016a}
D.~Torricelli, F.~Barroso, M.~Coscia, C.~Alessandro, F.~Lunardini,
  E.~Bravo~Esteban, and A.~D'Avella, ``{Muscle Synergies in Clinical Practice:
  Theoretical and Practical Implications},'' in {\em Emerging Therapies in
  Neurorehabilitation II} (J.~L. Pons, R.~Raya, and J.~Gonz{\'{a}}lez, eds.),
  vol.~10 of {\em Biosystems {\&} Biorobotics}, pp.~251--272, Cham: Springer
  International Publishing, 2016.

\bibitem{Rasool2016}
G.~Rasool, K.~Iqbal, N.~Bouaynaya, and G.~White, ``{Real-Time Task
  Discrimination for Myoelectric Control Employing Task-Specific Muscle
  Synergies.},'' {\em IEEE transactions on neural systems and rehabilitation
  engineering}, vol.~24, pp.~98--108, Jan. 2016.

\bibitem{Ison2015}
M.~Ison and P.~Artemiadis, ``{Proportional Myoelectric Control of Robots:
  Muscle Synergy Development Drives Performance Enhancement, Retainment, and
  Generalization},'' {\em IEEE Transactions on Robotics}, vol.~31,
  pp.~259--268, Apr. 2015.

\bibitem{Nazifi2017}
M.~M. Nazifi, H.~U. Yoon, K.~Beschorner, and P.~Hur, ``{Shared and
  Task-Specific Muscle Synergies during Normal Walking and Slipping},'' {\em
  Frontiers in Human Neuroscience}, vol.~11, pp.~1--14, Feb. 2017.

\bibitem{Martino2015}
G.~Martino, Y.~P. Ivanenko, A.~D'Avella, M.~Serrao, A.~Ranavolo, F.~Draicchio,
  G.~Cappellini, C.~Casali, and F.~Lacquaniti, ``{Neuromuscular adjustments of
  gait associated with unstable conditions},'' {\em Journal of
  Neurophysiology}, vol.~114, p.~jn.00029.2015, Sept. 2015.

\bibitem{Lee1999}
D.~D. Lee and H.~S. Seung, ``{Learning the parts of objects by non-negative
  matrix factorization.},'' {\em Nature}, vol.~401, pp.~788--91, Oct. 1999.

\bibitem{Jackson1991}
J.~E. Jackson, {\em {A User's Guide to Principal Components}}.
\newblock Wiley Series in Probability and Statistics, Hoboken, NJ, USA: John
  Wiley {\&} Sons, Inc., Mar. 1991.

\bibitem{Hyvarinen2000}
A.~Hyv{\"{a}}rinen and E.~Oja, ``{Independent component analysis: algorithms
  and applications},'' {\em Neural Networks}, vol.~13, pp.~411--430, June 2000.

\bibitem{Tresch2006}
M.~C. Tresch, V.~C.-K.~K. Cheung, and A.~D'Avella, ``{Matrix factorization
  algorithms for the identification of muscle synergies: evaluation on
  simulated and experimental data sets.},'' {\em Journal of neurophysiology},
  vol.~95, pp.~2199--2212, Apr. 2006.

\bibitem{Choi2011}
C.~Choi and J.~Kim, ``{Synergy matrices to estimate fluid wrist movements by
  surface electromyography},'' {\em Medical Engineering and Physics}, vol.~33,
  pp.~916--923, Oct. 2011.

\bibitem{DAvella2005}
A.~d'Avella and E.~Bizzi, ``{Shared and specific muscle synergies in natural
  motor behaviors.},'' {\em Proceedings of the National Academy of Sciences of
  the United States of America}, vol.~102, pp.~3076--81, Feb. 2005.

\bibitem{Cheung2005}
V.~C.-K.~K. Cheung, ``{Central and Sensory Contributions to the Activation and
  Organization of Muscle Synergies during Natural Motor Behaviors},'' {\em
  Journal of Neuroscience}, vol.~25, pp.~6419--6434, July 2005.

\bibitem{Torres-Oviedo2006}
G.~Torres-Oviedo, J.~M. Macpherson, and L.~H. Ting, ``{Muscle synergy
  organization is robust across a variety of postural perturbations.},'' {\em
  Journal of neurophysiology}, vol.~96, pp.~1530--1546, Jan. 2006.

\bibitem{Barroso2014}
F.~O. Barroso, D.~Torricelli, J.~C. Moreno, J.~Taylor, J.~Gomez-Soriano,
  E.~Bravo-Esteban, S.~Piazza, C.~Santos, and J.~L. Pons, ``{Shared muscle
  synergies in human walking and cycling},'' {\em Journal of Neurophysiology},
  vol.~112, no.~8, pp.~1984--1998, 2014.

\bibitem{Chvatal2011a}
S.~A. Chvatal, G.~Torres-Oviedo, S.~A. Safavynia, and L.~H. Ting, ``{Common
  muscle synergies for control of center of mass and force in nonstepping and
  stepping postural behaviors},'' {\em Journal of Neurophysiology}, vol.~106,
  pp.~999--1015, Aug. 2011.

\bibitem{Torres-Oviedo2010}
G.~Torres-Oviedo and L.~H. Ting, ``{Subject-Specific Muscle Synergies in Human
  Balance Control Are Consistent Across Different Biomechanical Contexts},''
  {\em Journal of Neurophysiology}, vol.~103, pp.~3084--3098, June 2010.

\bibitem{Atzori2014}
M.~Atzori, A.~Gijsberts, C.~Castellini, B.~Caputo, A.-G.~M. Hager, S.~Elsig,
  G.~Giatsidis, F.~Bassetto, and H.~M{\"{u}}ller, ``{Electromyography data for
  non-invasive naturally-controlled robotic hand prostheses.},'' {\em
  Scientific data}, vol.~1, p.~140053, Jan. 2014.

\bibitem{Atzori2015a}
M.~Atzori, A.~Gijsberts, I.~Kuzborskij, S.~Elsig, A.-G. Mittaz~Hager,
  O.~Deriaz, C.~Castellini, H.~Muller, and B.~Caputo, ``{Characterization of a
  Benchmark Database for Myoelectric Movement Classification},'' {\em IEEE
  Transactions on Neural Systems and Rehabilitation Engineering}, vol.~23,
  pp.~73--83, Jan. 2015.

\bibitem{Jiang2014b}
N.~Jiang, H.~Rehbaum, I.~Vujaklija, B.~Graimann, and D.~Farina, ``{Intuitive,
  online, simultaneous, and proportional myoelectric control over two
  degrees-of-freedom in upper limb amputees.},'' {\em IEEE transactions on
  neural systems and rehabilitation engineering}, vol.~22, no.~3, pp.~501--10,
  2014.

\bibitem{Ma2015a}
J.~Ma, N.~V. Thakor, and F.~Matsuno, ``{Hand and Wrist Movement Control of
  Myoelectric Prosthesis Based on Synergy},'' {\em IEEE Transactions on
  Human-Machine Systems}, vol.~45, pp.~74--83, Feb. 2015.

\bibitem{Lin2017}
C.~Lin, B.~Wang, N.~Jiang, and D.~Farina, ``{Robust extraction of basis
  functions for simultaneous and proportional myoelectric control via sparse
  non-negative matrix factorization},'' {\em Journal of Neural Engineering},
  vol.~15, p.~026017, Apr. 2018.

\bibitem{Debals2015}
O.~Debals and L.~De~Lathauwer, ``{Stochastic and Deterministic Tensorization
  for Blind Signal Separation},'' in {\em Latent Variable Analysis and Signal
  Separation}, pp.~3--13, Springer, Cham, 2015.

\bibitem{Comon2014}
P.~Comon, ``{Tensors : A brief introduction},'' {\em IEEE Signal Processing
  Magazine}, vol.~31, pp.~44--53, May 2014.

\bibitem{Kolda2008c}
T.~G. Kolda and B.~W. Bader, ``{Tensor Decompositions and Applications},'' {\em
  SIAM Review}, vol.~51, pp.~455--500, Aug. 2008.

\bibitem{Bro2003}
R.~Bro and H.~A.~L. Kiers, ``{A new efficient method for determining the number
  of components in PARAFAC models},'' {\em Journal of Chemometrics}, vol.~17,
  pp.~274--286, June 2003.

\bibitem{Zhou2012}
G.~Zhou and A.~Cichocki, ``{Fast and unique Tucker decompositions via multiway
  blind source separation},'' {\em Bulletin of the Polish Academy of Sciences:
  Technical Sciences}, vol.~60, pp.~389--405, Jan. 2012.

\bibitem{Harshman1994}
R.~A. Harshman and M.~E. Lundy, ``{PARAFAC: Parallel factor analysis},'' {\em
  Computational Statistics {\&} Data Analysis}, vol.~18, pp.~39--72, Aug. 1994.

\bibitem{Sands1980}
R.~Sands and F.~W. Young, ``{Component models for three-way data: An
  alternating least squares algorithm with optimal scaling features},'' {\em
  Psychometrika}, vol.~45, pp.~39--67, Mar. 1980.

\bibitem{Comon2009}
P.~Comon, X.~Luciani, and A.~L.~F. de~Almeida, ``{Tensor decompositions,
  alternating least squares and other tales},'' {\em Journal of Chemometrics},
  vol.~23, pp.~393--405, July 2009.

\bibitem{Cichocki2009b}
A.~Cichocki, R.~Zdunek, A.~H. Phan, and S.-I. Amari, {\em {Nonnegative Matrix
  and Tensor Factorizations: Applications to Exploratory Multi-Way Data
  Analysis and Blind Source Separation}}, vol.~1.
\newblock John Wiley {\&} Sons, 2009.

\bibitem{Liu2008}
S.~Liu and G.~Trenkler, ``{Hadamard, Khatri-Rao, Kronecker and other matrix
  products},'' {\em Int. J. Inf. Syst. Sci}, vol.~4, no.~1, pp.~160--177, 2008.

\bibitem{Andersson2000}
C.~A. Andersson and R.~Bro, ``{The N-way Toolbox for MATLAB},'' {\em
  Chemometrics and Intelligent Laboratory Systems}, vol.~52, pp.~1--4, Aug.
  2000.

\bibitem{Nie2009}
F.~Nie, S.~Xiang, Y.~Song, and C.~Zhang, ``{Extracting the optimal
  dimensionality for local tensor discriminant analysis},'' {\em Pattern
  Recognition}, vol.~42, no.~1, pp.~105--114, 2009.

\bibitem{Ebied2018}
A.~Ebied, E.~Kinney-Lang, L.~Spyrou, and J.~Escudero, ``{Evaluation of matrix
  factorisation approaches for muscle synergy extraction},'' {\em Medical
  Engineering {\&} Physics}, vol.~57, pp.~51--60, July 2018.

\bibitem{Muceli2012c}
S.~Muceli and D.~Farina, ``{Simultaneous and proportional estimation of hand
  kinematics from EMG during mirrored movements at multiple
  degrees-of-freedom},'' {\em IEEE Transactions on Neural Systems and
  Rehabilitation Engineering}, vol.~20, pp.~371--378, May 2012.

\bibitem{Jiang2009}
N.~Jiang, K.~B. Englehart, and P.~a. Parker, ``{Extracting simultaneous and
  proportional neural control information for multiple-dof prostheses from the
  surface electromyographic signal},'' {\em IEEE Transactions on Biomedical
  Engineering}, vol.~56, pp.~1070--1080, Apr. 2009.

\bibitem{Ikeda2000}
S.~Ikeda and K.~Toyama, ``{Independent component analysis for noisy data - MEG
  data analysis},'' {\em Neural Networks}, vol.~13, no.~10, pp.~1063--1074,
  2000.

\bibitem{Chvatal2013a}
S.~A. Chvatal and L.~H. Ting, ``{Common muscle synergies for balance and
  walking.},'' {\em Frontiers in computational neuroscience}, vol.~7, p.~48,
  Jan. 2013.

\bibitem{Devarajan2008}
K.~Devarajan, ``{Nonnegative matrix factorization: an analytical and
  interpretive tool in computational biology.},'' {\em PLoS computational
  biology}, vol.~4, p.~e1000029, Jan. 2008.

\bibitem{Semprini2017}
M.~Semprini, A.~V. Cuppone, I.~Delis, V.~Squeri, S.~Panzeri, and J.~Konczak,
  ``{Biofeedback Signals for Robotic Rehabilitation: Assessment of Wrist Muscle
  Activation Patterns in Healthy Humans},'' {\em IEEE Transactions on Neural
  Systems and Rehabilitation Engineering}, vol.~25, pp.~883--892, July 2017.

\bibitem{Mrup2011b}
M.~M{\o}rup, ``{Applications of tensor (multiway array) factorizations and
  decompositions in data mining},'' {\em Wiley Interdisciplinary Reviews: Data
  Mining and Knowledge Discovery}, vol.~1, pp.~24--40, Jan. 2011.

\bibitem{Jiang2014a}
N.~Jiang, T.~Lorrain, and D.~Farina, ``{A state-based, proportional myoelectric
  control method: online validation and comparison with the clinical
  state-of-the-art.},'' {\em Journal of neuroengineering and rehabilitation},
  vol.~11, p.~110, July 2014.

\end{thebibliography}

\EOD

\end{document}